\documentclass[superscriptaddress,showpacs,nofootinbib,notitlepage,preprintnumbers,secnumarabic,amssymb, nobibnotes,aps, prd,twocolumn,10pt]{revtex4-1}
\usepackage[utf8]{inputenc}
\usepackage{graphicx}
\usepackage{latexsym,amsmath,amssymb,amsthm,lmodern,float,url,bm}
\usepackage{natbib}
\usepackage{color}
\usepackage{microtype}
\usepackage{import}
\usepackage{bbold}
\usepackage[plain]{fancyref}
\usepackage{varioref}
\usepackage{slashed}
\usepackage{multirow}
\usepackage{tikz}
\usepackage{physics}
\usepackage{comment}
\usepackage[]{units}
\usetikzlibrary{shapes}
\usetikzlibrary{positioning}

\usepackage{silence}
\usepackage[colorlinks=true,backref=false, linktocpage=true,
citecolor=blue,urlcolor=blue,linkcolor=blue,pdfpagemode=UseOutlines]{hyperref}
\hypersetup{%
  bookmarksnumbered=true,
  pdftitle = {},
  pdfsubject = {},
  pdfauthor = {},
  pdfkeywords = {}
}

\newcommand{\GeV}{\,\text{GeV}}
\newcommand{\MeV}{\,\text{MeV}}

\allowdisplaybreaks

\begin{document}
\preprint{EOS-2025-02, ZU-TH 04/25}
\title{A model-independent parameterization of \texorpdfstring{$B\rightarrow\pi\pi\ell\nu$}{B→ππlν} decays}

\author{Florian Herren}
\affiliation{Physics Institute, Universität Zürich, Winterthurerstrasse 190, CH-8057 Zürich, Switzerland}

\author{Bastian Kubis}
\affiliation{Helmholtz-Institut für Strahlen- und Kernphysik (Theorie) and Bethe Center for Theoretical Physics,
Universität Bonn, 53115 Bonn, Germany}

\author{Raynette van Tonder}
\affiliation{Institut für Experimentelle Teilchenphysik, Karlsruhe Institute of Technology (KIT), D-76131 Karlsruhe, Germany}

\date{\today}

\begin{abstract}
We introduce a novel parameterization of $B\rightarrow\pi\pi\ell\nu$ form factors relying on partial-wave decompositions and series expansions in suitable variables. We bound the expansion coefficients through unitarity and include left-hand cut contributions using established dispersive methods. The two-hadron lineshapes are treated in a model-independent manner using Omnès functions, thus allowing for a data-driven determination of the expansion parameters. We study the underlying composition of the di-pion system in $B\rightarrow\pi\pi\ell\nu$ decays through fits to differential spectra of $B^{+} \rightarrow \pi^{+}\pi^{-} \ell^+ \nu$ measured by the Belle experiment. In contrast to previous works, we are able to study the full phase-space and are not limited to certain kinematic regions. As a consequence, we extract branching fractions for the different partial waves of the di-pion system. We find:
\begin{align*}
    \mathcal{B}(B^+\rightarrow (\pi^+ \pi^-)_S \ell^+ \nu) &= 2.2^{+1.4}_{-1.0}\times 10^{-5}\,,\nonumber\\
    \mathcal{B}(B^+\rightarrow (\pi^+ \pi^-)_P \ell^+\nu) &= 19.6^{+2.8}_{-2.7}\times 10^{-5}\,,\nonumber\\
    \mathcal{B}(B^+\rightarrow (\pi^+ \pi^-)_D \ell^+ \nu) &= 3.5^{+1.3}_{-1.1}\times 10^{-5}\,.
\end{align*}
In addition, we derive predictions for the thus far unobserved $B^+\rightarrow \pi^0\pi^0\ell^+\nu$ decay and obtain a sizeable branching fraction of $\mathcal{B}(B^+\rightarrow \pi^0\pi^0\ell^+\nu) = 2.9^{+0.9}_{-0.7}\times 10^{-5}$.
\end{abstract}

\maketitle

\section{Introduction}
Precise determinations of the Cabibbo--Kobayashi--Maskawa (CKM) matrix elements allow for potent tests of the Standard Model (SM) by overconstraining the CKM unitarity triangle in global fits~\cite{Charles:2004jd,HeavyFlavorAveragingGroupHFLAV:2024ctg,UTfit:2022hsi}. A well-established strategy to determine the magnitude of the matrix element $|V_{ub}|$ is through measurements of semileptonic $B$ meson decays, which allow for greater theoretical control than decays involving purely hadronic final states. Determinations of $|V_{ub}|$ are extracted by employing two complementary approaches: the exclusive approach focuses on the reconstruction of specific decay modes, while the inclusive approach aims to measure the sum of all possible final states entailing the same quark-level transition. Current world averages of $|V_{ub}|$ from exclusive and inclusive determinations exhibit a disagreement of approximately 3 standard deviations, a longstanding and unresolved puzzle to date.

\begin{figure}
    \centering
    \includegraphics[width=\linewidth]{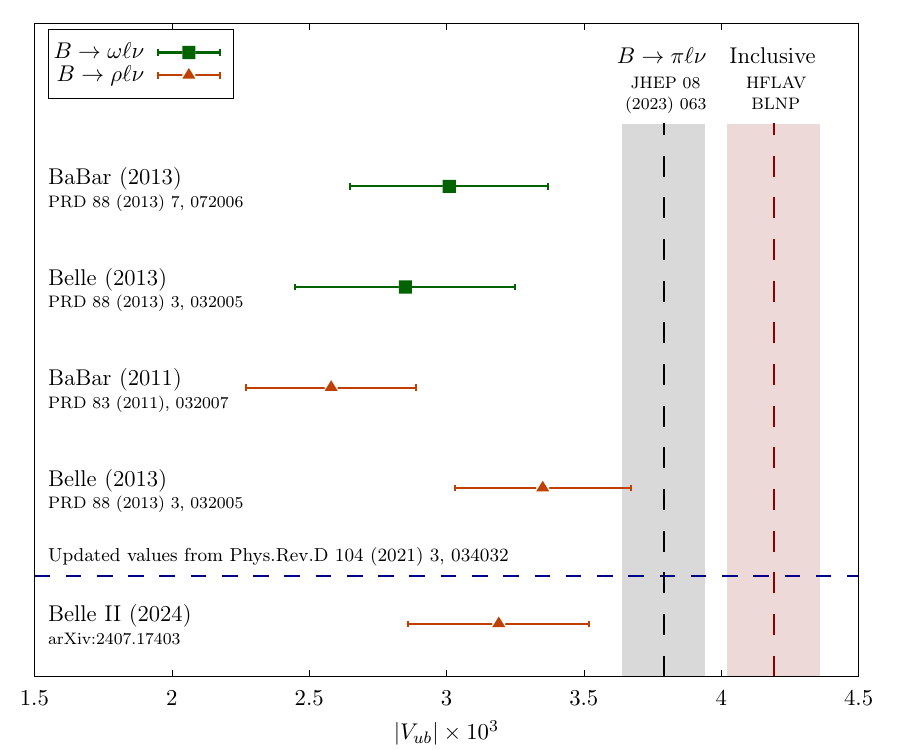}
    \caption{The extracted $|V_{ub}|$ values for $B \rightarrow \omega \ell \nu$ and $B \rightarrow \rho \ell \nu$ from BaBar~\cite{BaBar:2010efp,BaBar:2013pls}, Belle~\cite{Belle:2013hlo}, and Belle II~\cite{Belle-II:2024xwh}, compared to the values extracted in a global fit ot $B\rightarrow \pi\ell\nu$ data in Ref.~\cite{Leljak:2023gna} and the average of inclusive determinations~\cite{HeavyFlavorAveragingGroupHFLAV:2024ctg}. The $B \rightarrow \omega \ell \nu$ and $B \rightarrow \rho \ell \nu$ values for BaBar and Belle have been updated with new form factor input in Ref.~\cite{Bernlochner:2021rel}.}
    \label{fig:vub_status}
\end{figure}

While the most precise exclusive determinations of $|V_{ub}|$ are extracted from measurements of $B \rightarrow \pi \ell \nu$ decays~\cite{HeavyFlavorAveragingGroupHFLAV:2024ctg}, measurements of $B \rightarrow \omega \ell \nu$ and $B \rightarrow \rho \ell \nu$ have also been performed. Here, $\omega$ and $\rho$ refer to the $\omega(782)$ and $\rho(770)$, respectively. Interestingly, extractions of $|V_{ub}|$ using $B \rightarrow \omega \ell \nu$ decays are compatible with each other, but are systematically lower than determinations from $B \rightarrow \pi \ell \nu$~\cite{BaBar:2012dvs,Belle:2013hlo}. The situation is further complicated when considering $B \rightarrow \rho \ell \nu$ decays where the two most precise measurements are in significant tension with each other~\cite{BaBar:2010efp,Belle:2013hlo}. Additionally, the result reported by Belle is compatible with the world average of $|V_{ub}|$ from the $B \rightarrow \pi \ell \nu$ mode, while BaBar quoted a lower value. Recently, Belle II reported a tagged analysis of $B^+ \rightarrow \rho^0 \ell^+ \nu$~\cite{Belle-II:2022fsw} as well as a simultaneous analysis of $B^{0} \rightarrow \pi^{-} \ell^{+} \nu$ and $B^{+} \rightarrow \rho^{0} \ell^{+} \nu$ decays using an untagged reconstruction method~\cite{Belle-II:2024xwh}. The branching fraction extracted in the former is compatible with the one obtained by Belle, but not with the BaBar measurement. The latter determined $|V_{ub}|$ from the $B^+ \rightarrow \rho^0 \ell^+ \nu$ mode that is compatible with the measurements by BaBar and Belle. 
Figure~\ref{fig:vub_status} shows the current status of extracted $|V_{ub}|$ values for $B \rightarrow \omega \ell \nu$ and $B \rightarrow \rho \ell \nu$ from different experiments, updated in Ref.~\cite{Bernlochner:2021rel} to the more recent form factor calculation of Ref.~\cite{Bharucha:2015bzk}.

Charmless semileptonic decays are typically modeled as a mixture of specific exclusive modes and non-resonant contributions. Various different approaches are employed to combine simulated decays of known resonances, namely $B \rightarrow \{\pi, \omega, \rho, \eta,  \eta^{\prime}\} \ell \nu$, with scaled predictions of the total inclusive $B\rightarrow X_{u}\ell \nu$ decay rate. Pythia~\cite{Sjostrand:1993yb} is then generally used to hadronize the inclusive spectrum into various hadronic final states. Both exclusive and inclusive experimental measurements rely on Monte Carlo (MC) simulations to either subtract or include non-resonant $B\rightarrow X_{u}\ell \nu$ processes, the size of which is highly dependent on the underlying theoretical description or MC methodology. Studies that make use of a “hybrid” method, originally proposed by Ref.~\cite{Ramirez:1989yk}, to combine exclusive decay modes with inclusive predictions report a different estimation of the non-resonant contribution compared to studies that make use of alternative methods~\cite{Belle:2009pop}. The modeling of this inclusive non-resonant component becomes, in turn, one of the leading sources of systematic error for not only studies of exclusive modes such as $B \rightarrow \rho \ell \nu$~\cite{BaBar:2010efp,Belle:2013hlo,Belle-II:2022fsw,Belle-II:2024xwh}, $B^{+}\rightarrow \mu^{+}\nu$~\cite{Belle:2019iji} and $B^+ \to \gamma\ell^{\,+} \nu$~\cite{Belle:2018jqd}, but also inclusive determinations of $|V_{ub}|$~\cite{Belle:2021eni}. In addition, inclusive analyses measuring kinematic distributions of $B \rightarrow X_{c} \ell \nu$ decays usually reconstruct $B \rightarrow X \ell \nu$ decays and subtract the significantly smaller $B \rightarrow X_{u} \ell \nu$ component, treated as a background process, based on estimations from simulation. As a result, this strategy leads to a non-negligible modeling uncertainty in recent measurements of $B \rightarrow X_{c} \ell \nu$ kinematic spectra~\cite{Belle:2021idw,Belle-II:2022evt}.

To improve future measurements of the $\rho^{0}$ final state and investigate further unflavored resonances decaying to a charged-pion pair, we investigate the four-body semileptonic decay $B^{+} \rightarrow \pi^{+}\pi^{-} \ell^+ \nu$. This channel is of particular interest, since the $\pi^{+}\pi^{-}$ system potentially comprises narrow resonances, broad states with nontrivial lineshapes as well as interference patterns. Differential kinematic spectra of the $B^{+} \rightarrow \pi^{+}\pi^{-} \ell^+ \nu$ decay have been measured by the Belle Collaboration in Ref.~\cite{Belle:2020xgu}. By performing a two-dimensional analysis of the partial branching fractions as a function of the di-pion invariant mass, $M_{\pi \pi}$, and the four-momentum transfer squared, $q^{2}$, this measurement allows for a unique probe of the composition of the $\pi^{+}\pi^{-}$ system. Using the spectra provided by this measurement, we study the underlying composition of the di-pion system by employing model-independent information to explicitly describe the lineshapes of different partial waves.

By virtue of Watson's theorem~\cite{Watson:1952ji}, we are able to harness the high precision obtained on the $\pi\pi$ scattering phase shifts by means of Roy equations~\cite{Roy:1971tc} and available $\pi\pi$ scattering data in Refs.~\cite{Ananthanarayan:2000ht,Garcia-Martin:2011iqs,Caprini:2011ky}. These analyses were further refined by including data on $e^+ e^- \rightarrow \pi^+ \pi^-$ for the P-wave~\cite{Colangelo:2018mtw} and differential decay rates in $B_{(s)}\rightarrow J/\Psi\pi\pi$ decays for the S-wave~\cite{Daub:2015xja,Ropertz:2018stk}. A previous attempt to develop a theoretical description of $B\rightarrow\pi\pi\ell\nu$ decays based on Heavy-Meson Chiral Perturbation Theory that includes the available information on the lineshapes, as well as left-hand cuts, was limited to the large-$q^2$ region of the phase space~\cite{Kang:2013jaa} and thus cannot be applied directly to the Belle data. Consequently, we aim to extend the parameterization of Ref.~\cite{Gustafson:2023lrz}, developed to study $B\rightarrow D\pi\ell\nu$ decays, that incorporates unitarity bounds on the relevant form factors and is not limited in the $q^2$-range. However, left-hand cuts and inelasticities that are relevant for $B\rightarrow\pi\pi\ell\nu$ decays are not accounted for, which we will resolve in this work.

The remainder of this paper is structured as follows. We introduce the five-fold differential decay rate of $B\rightarrow\pi\pi\ell\nu$ decays in Sec.~\ref{sec::Setup}. In Sec.~\ref{sec::Formalism} we present a novel form factor decomposition with the correct analytic structure for a three-hadron form factor and derive a parameterization of the form factors bounded by unitarity.
This parameterization requires a model-independent treatment of the di-pion invariant-mass spectrum, which is discussed in Sec.~\ref{sec::SPD}. With this parameterization at hand, we perform a fit to the Belle measurement of Ref.~\cite{Belle:2020xgu} and discuss the results in Sec.~\ref{sec::Belle}. Finally, we conclude with a discussion of the implications of our findings and an outlook on possible extensions of our work in Sec.~\ref{sec::Outlook}.

\section{The \texorpdfstring{$B\rightarrow \pi\pi\ell\nu$}{B→ππlν} decay rate}\label{sec::Setup}
\subsection{Kinematics}
The decay $B(p_B) \rightarrow \pi(p_1)\pi(p_2)\ell(p_\ell)\nu(p_\nu)$ is characterized by five independent kinematic quantities: two invariant masses, $q^2 = (p_\ell + p_\nu)^2$, $s = (p_1+p_2)^2$, the azimuthal angle between the di-lepton and di-pion decay planes $\chi$, as well as $\theta_\ell$ and $\theta_\pi$, the polar angles of the lepton and the pion in the di-lepton and di-pion restframes.

To relate the angles to scalar products between the four-momenta, we introduce the differences
\begin{align}
    \delta_{12} = p_1 - p_2\,,\quad\delta_{\ell\nu} = p_\ell - p_\nu
\end{align}
as well as the projectors
\begin{align}
    P^{(q)}_{\mu\nu} = -g_{\mu\nu} + \frac{q_\mu q_\nu}{q^2}\,,~ P^{(12)}_{\mu\nu} = -g_{\mu\nu} + \frac{p_{12,\mu} p_{12,\nu}}{s}\,,
\end{align}
where $p_{12}^\mu = p_1^\mu + p_2^\mu$\,. Computing products between momenta and projectors, we obtain
\begin{align}
    q^2 &\left(p_{12}\cdot P^{(q)}\cdot p_{12}\right) = s\left(q\cdot P^{(12)}\cdot q\right) = \frac{\lambda_{B\ell}}{4}\,, \notag\\
    &\left(\delta_{\ell\nu}\cdot P^{(q)}\cdot \delta_{\ell\nu}\right) = q^2\beta^2_\ell\,,\notag\\&\left(\delta_{12}\cdot P^{(12)}\cdot \delta_{12}\right) = s\beta^2_\pi\,, \notag\\
    &\left(p_{12}\cdot P^{(q)}\cdot \delta_{\ell\nu}\right) = \frac{\kappa_{\ell\nu}}{2}\cos\theta_\ell\,, \notag\\
    &\left(q\cdot P^{(12)}\cdot \delta_{12}\right) = \frac{\kappa_{12}}{2}\cos\theta_\pi\,,
\end{align}
where $\lambda_{B\ell} = \lambda(M_B^2,q^2,s)$ with the Källén function
\begin{align}
    \lambda(x,y,z) = x^2 + y^2 + z^2 - 2(xy+yz+zx)\,,
\end{align}
while
\begin{align}
    \beta_\ell &= \frac{\sqrt{\lambda(q^2,m_\ell^2,0)}}{q^2} = 1-\frac{m_\ell^2}{q^2}\,,\nonumber\\
    \beta_\pi &= \frac{\sqrt{\lambda(s,M_\pi^2,M_\pi^2)}}{s} = \sqrt{1-\frac{4M_\pi^2}{s}}\,.
\end{align}
Finally, $\kappa_{\ell\nu} = \beta_\ell\sqrt{\lambda_{B\ell}}$ and $\kappa_{12} = \beta_\pi\sqrt{\lambda_{B\ell}}$.

Furthermore, one additional vector orthogonal to both $q^\mu$ and $p_{12}^\mu$ is required for the form factor decomposition.
We take
\begin{align}
    T^{(12)}_\mu = P^{(12)}_{\mu\nu}\delta_{12}^\nu - \frac{\left(q\cdot P^{(12)}\cdot \delta_{12}\right)}{\left(q\cdot P^{(12)}\cdot q\right)}P^{(12)}_{\mu\nu}q^\nu\,,
\end{align}
which fulfills
\begin{align}
    \delta_{12}^\mu T^{(12)}_\mu = -(T^{(12)})^2 = s\frac{\kappa^2_{12}}{\lambda_{B\ell}}\sin^2\theta_\pi\,.
\end{align}
This scalar product is closely related to contractions of the Levi-Civita tensor with all three meson momenta:
\begin{align}
    (i\epsilon_{\mu\nu\rho\sigma}p_1^\nu p_2^\rho p_B^\sigma)^2 = s\frac{\kappa^2_{12}}{16}\sin^2\theta_\pi\,.
\end{align}

Finally, for the computation of the decay rate, the quantities
\begin{align}
    \delta^\mu_{\ell\nu} T^{(12)}_\mu = \sqrt{s}\sqrt{q^2}\frac{\kappa_{\ell\nu}\kappa_{12}}{\lambda_{B\ell}}\sin\theta_\pi\sin\theta_\ell\cos\chi
\end{align}
and
\begin{align}
    i\epsilon_{\mu\nu\rho\sigma}\delta^\mu_{\ell\nu}q^\nu\delta_{12}^\rho p_{12}^\sigma = i\sqrt{s}\sqrt{q^2}\frac{\kappa_{\ell\nu}\kappa_{12}}{2\sqrt{\lambda_{B\ell}}}\sin\theta_\ell\sin\theta_\pi\sin\chi
\end{align}
are required.

\subsection{Form factors}
With the quantities introduced in the previous section at hand, we can write down a fully general form factor decomposition for $B\rightarrow\pi\pi\ell\nu$ decays:
\begin{align}
    \left\langle \pi^j(p_1)\pi^k(p_2)|V_\mu|B(p_B)\right\rangle &= i\epsilon_{\mu\nu\rho\sigma}p_B^\nu p_1^\rho p_2^\sigma \,g^{(jk)}(s,t,u)\,,\nonumber\\
    \left\langle \pi^j(p_1)\pi^k(p_2)|A_\mu|B(p_B)\right\rangle &= T_\mu^{(12)}\,f^{(jk)}(s,t,u)\nonumber\\ &+ P^{(q)}_{\mu\nu}p_{12}^\nu\,\mathcal{F}^{(jk)}_1(s,t,u)\nonumber\\ &+ \frac{q^\mu}{q^2}\,\mathcal{F}^{(jk)}_2(s,t,u)\,.
\end{align}
Here, the labels $j, k \in \{0,+,-\}$ denote the charges of the pions, which are relevant to determine the respective isospin relations later on. To simplify the notation, we write
\begin{align}
    \mathcal{M}^\mu_{jk} = \left\langle \pi^j(p_1)\pi^k(p_2)|V^\mu-A^\mu|B(p_B)\right\rangle\,.
\end{align}

The form factors introduced here differ from those of Ref.~\cite{Faller:2013dwa} by kinematic factors that can become singular. Each of the form factors depends on three independent kinematic variables, making a model-independent description significantly more cumbersome than in the case of $1\rightarrow 1$ transitions.

Various techniques have been applied to parameterize form factors for $B\rightarrow\pi\pi$ transitions or similar $1\rightarrow 2$ or $0\rightarrow 3$ form factors. For the case of $\gamma^\ast\rightarrow 3\pi$ dispersive parameterizations exist~\cite{Hoferichter:2018dmo,Hoferichter:2023bjm}, while $K\rightarrow \pi\pi\ell\nu$ decays have been studied using reconstruction theorems to obtain the full $s$-, $t$-, and $u$-dependence~\cite{Colangelo:2015kha}. For phenomenological studies of $B\rightarrow D\pi\ell\nu$ decays~\cite{Gustafson:2023lrz} and light-cone sum rule (LCSR) calculations of $B\rightarrow K\pi\ell\ell$~\cite{Descotes-Genon:2019bud,Descotes-Genon:2023ukb} or $B\rightarrow \pi\pi\ell\nu$~\cite{Hambrock:2015aor,Cheng:2017smj,Cheng:2017sfk,Cheng:2025hxe} decays a partial-wave expansion and subsequent factorization of the $q^2$- and $s$-dependence is employed. The dispersive treatment of Ref.~\cite{Kang:2013jaa} and the Quantum Chromodynamics (QCD) factorization-based calculations of Refs.~\cite{Boer:2016iez,Feldmann:2018kqr} also include crossed-channel $B^\ast$ contributions.

In Sec.~\ref{sec::Formalism} we introduce a novel parameterization, combining the strength of the dispersive representations introduced in Refs.~\cite{Kang:2013jaa,Colangelo:2015kha} with the unitarity bounds derived in Ref.~\cite{Gustafson:2023lrz}.

\subsection{Five-fold differential decay rate}
The five-fold differential decay rate is given by
\begin{align}
    \frac{\mathrm{d}^5\Gamma_{B\rightarrow\pi^j\pi^k\ell\nu}}{\mathrm{d}q^2\,\mathrm{d}s\,\mathrm{d}\cos\theta_\pi\,\mathrm{d}\cos\theta_\ell\,\mathrm{d}\chi} = K_{jk}\kappa_{12}\beta_\ell\mathcal{M}^\mu_{jk}\mathcal{M}^{\ast,\nu}_{jk} L_{\mu\nu}\,,
\end{align}
where the constant factor $K_{jk}$ and the leptonic tensor $L^{\mu\nu}$ are given by
\begin{align}
    K_{jk} &= \frac{G_F^2|V_{ub}|^2}{4^7\pi^6 M_B^3 n_{jk}}\, ,\nonumber\\
    L^{\mu\nu} &= P^{(q),\mu\nu}(q^2-m_\ell^2) + \frac{q^\mu q^\nu}{q^2}m_\ell^2\nonumber\\ &- \delta_{\ell\nu}^\mu\delta_{\ell\nu}^\nu - i \epsilon^{\mu\nu\rho\sigma}q_\rho\delta_{\ell\nu,\sigma}\,.
\end{align}
Here, $G_F$ is Fermi's constant and the symmetrization factor $n_{jk}$ is $2$ for $j,k = 0$ and $1$ otherwise. Evaluating the product between hadronic and leptonic tensor yields
\begin{align}
    &\mathcal{M}^\mu_{jk}\mathcal{M}^{\ast,\nu}_{jk} L_{\mu\nu} = q^2\beta_\ell\Big(M^{(jk)}_1 + M^{(jk)}_2 \cos 2\theta_\ell \nonumber\\ &+ M^{(jk)}_3\sin^2\theta_\ell\cos 2\chi + M^{(jk)}_4 \sin 2\theta_\ell\cos\chi\nonumber\\
    & + M^{(jk)}_5\sin\theta_\ell\cos\chi  + M^{(jk)}_6 \cos\theta_\ell+ M^{(jk)}_7\sin\theta_\ell\sin\chi \nonumber\\
    &+ M^{(jk)}_8 \sin 2\theta_\ell \sin\chi + M^{(jk)}_9 \sin^2\theta_\ell\sin 2\chi\Big)\,,
\end{align}
where the $M^{(jk)}_i$ are combinations of kinematic factors and form factors:
\begin{align}
    M^{(jk)}_1 &= \Big(1-\frac{\beta_\ell}{4}\Big)\left(|\mathcal{A}^{(jk)}_\parallel|^2+|\mathcal{A}^{(jk)}_\perp|^2\right)\nonumber\\ &+ \Big(1-\frac{\beta_\ell}{2}\Big)|\mathcal{A}^{(jk)}_0|^2 + \frac{m_\ell^2}{q^2}|\mathcal{A}^{(jk)}_t|^2\,,\nonumber\\
    M^{(jk)}_2 &= \beta_\ell\left[\frac{1}{4}\left(|\mathcal{A}^{(jk)}_\parallel|^2+|\mathcal{A}^{(jk)}_\perp|^2\right) - \frac{1}{2}|\mathcal{A}^{(jk)}_0|^2\right]\,,\nonumber\\
    M^{(jk)}_3 &= \frac{1}{2}\beta_\ell\left[|\mathcal{A}^{(jk)}_\perp|^2 - |\mathcal{A}^{(jk)}_\parallel|^2\right]\,,\nonumber\\
    M^{(jk)}_4 &= \beta_\ell\, \mathrm{Re}(\mathcal{A}^{(jk)}_0 \mathcal{A}^{(jk),\ast}_\parallel)\,,\nonumber\\
    M^{(jk)}_5 &= -2 \left[\mathrm{Re}(\mathcal{A}^{(jk)}_0 \mathcal{A}^{(jk),\ast}_\perp) + \frac{m_\ell^2}{q^2}\mathrm{Re}(\mathcal{A}^{(jk)}_t \mathcal{A}^{(jk),\ast}_\parallel) \right],\nonumber\\
    M^{(jk)}_6 &= 2 \left[\frac{m^2_\ell}{q^2}\mathrm{Re}(\mathcal{A}^{(jk)}_t \mathcal{A}^{(jk),\ast}_0) - \mathrm{Re}(\mathcal{A}^{(jk)}_\perp \mathcal{A}^{(jk),\ast}_\parallel)\right],\nonumber\\
    M^{(jk)}_7 &= -2 \left[\mathrm{Im}(\mathcal{A}^{(jk)}_0 \mathcal{A}^{(jk),\ast}_\parallel) -\frac{m_\ell^2}{q^2}\mathrm{Im}(\mathcal{A}^{(jk)}_t \mathcal{A}^{(jk),\ast}_\perp)\right],\nonumber\\
    M^{(jk)}_8 &= -\beta_\ell\, \mathrm{Im}(\mathcal{A}^{(jk)}_0\mathcal{A}^{(jk),\ast}_\perp)\,,\nonumber\\
    M^{(jk)}_9 &= \beta_\ell\, \mathrm{Im}(\mathcal{A}^{(jk)}_\perp\mathcal{A}^{(jk),\ast}_\parallel)\,,
\end{align}
where
\begin{align}
    \mathcal{A}^{(jk)}_\perp &= \frac{\sqrt{s}\kappa_{12}}{4}\sin\theta_\pi g^{(jk)}\,, & 
    \mathcal{A}^{(jk)}_\parallel &= \frac{\sqrt{s}\kappa_{12}}{\sqrt{\lambda_{B\ell}}}\sin\theta_\pi f^{(jk)}\,,\nonumber\\
    \mathcal{A}^{(jk)}_0 &= \frac{\sqrt{\lambda_{B\ell}}}{2\sqrt{q^2}} \mathcal{F}^{(jk)}_1\,, & 
    \mathcal{A}^{(jk)}_t &= \sqrt{q^2} \mathcal{F}^{(jk)}_2\,.
\end{align}

Integrating over $\cos\theta_\ell$ and $\chi$ leaves us with the triple differential decay rate
\begin{align}
    \frac{\mathrm{d}^3\Gamma_{B\rightarrow\pi^j\pi^k\ell\nu}}{\mathrm{d}q^2\,\mathrm{d}s\,\mathrm{d}\cos\theta_\pi} &= \frac{G_F^2|V_{ub}|^2}{M_B^3 n_{jk}}\frac{\kappa_{12}\beta_\ell}{4^6\pi^5}\left[M^{(jk)}_1 - \frac{M^{(jk)}_2}{3}\right]\,,
\end{align}
which cannot be further simplified without a parameterization of the form factors.

\section{Form factor parameterization}\label{sec::Formalism}
The form factors introduced in Sec.~\ref{sec::Setup} depend on three independent variables: $s$, $t = (p_B - p_1)^2$, and $u = (p_B - p_2)^2$. Consequently, they exhibit a complex analytic structure. As the derivation of a model-independent parameterization for such form factors is lengthy, we split this section into several parts. First, we introduce the main idea that would be valid in the absence of branch cuts induced by $t$- or $u$-channel $B\pi$ interactions and ignore the underlying isospin structure. Next, we introduce single-variable functions and discuss their general isospin decomposition, relevant to obtain the correct relations between $B^+\rightarrow \pi^+ \pi^- \ell^+\nu$, $B^+\rightarrow \pi^0 \pi^0 \ell^+\nu$, and $B^0\rightarrow \pi^- \pi^0 \ell^+\nu$ form factors. Afterwards, we derive reconstruction theorems relating the single-variable functions to the original form factors, depending on $s$, $t$, and $u$. Finally, we derive the unitarity bounds and a parameterization for the full system of single-variable functions.

\subsection{Main idea}
For the semileptonic decays under study, we are mainly interested in the analytic properties in $q^2$ and $s$, while expressing $t$ and $u$ through more convenient kinematic variables. To this end, we introduce the helicity angle of the positively charged pion in the di-pion restframe
\begin{align}
    \cos\theta_\pi = \frac{t - u}{\beta_\pi \sqrt{\lambda_{B\ell}}}
\end{align}
and eliminate the remaining dependence on $t + u$ in terms of $s$, $q^2$ and the particle masses.

To discuss the analytic structure and overall kinematic factors, we study the form factors in $2\rightarrow 2$ scattering kinematics, i.e., we consider the process $J B \rightarrow \pi \pi$. First, we perform a partial-wave expansion of the form factors in $\cos\theta_\pi$. This step is crucial to disentangle the contribution of different resonances in the $\pi\pi$ spectrum and separate isovector and isoscalar configurations, as even partial waves can only contain isoscalar $\pi\pi$ configurations, while the odd ones contain only the isovector ones. The exact form of the partial-wave expansion depends on the form factor under question: angular momentum conservation dictates that $\mathcal{F}_1$ and $\mathcal{F}_2$ are expanded in simple Legendre polynomials of $\cos\theta_\pi$, while $f$ and $g$ are expanded in terms of their derivatives~\cite{Jacob:1959at}:
\begin{align}
    \mathcal{F}_1(q^2,s,\cos\theta_\pi) &= \sum_{l = 0} P_l(\cos\theta_\pi)\mathcal{F}^{(l)}_{1}(q^2,s)\,,\nonumber\\
    \mathcal{F}_2(q^2,s,\cos\theta_\pi) &= \sum_{l = 0} P_l(\cos\theta_\pi)\mathcal{F}^{(l)}_{2}(q^2,s)\,,\nonumber\\
    f(q^2,s,\cos\theta_\pi) &= \sum_{l = 1} P^\prime_l(\cos\theta_\pi)f^{(l)}(q^2,s)\,,\nonumber\\
    g(q^2,s,\cos\theta_\pi) &= \sum_{l = 1} P^\prime_l(\cos\theta_\pi)g^{(l)}(q^2,s)\,.\label{eq::ffs_ang_simple}
\end{align}
The partial-wave amplitudes $F^{(l)}(q^2,s)$ all share similar properties. First, for $q^2 < q^2_- \equiv (M_B - 2M_\pi)^2$ they are real in the region $0 < s < 4M_\pi^2$ and, by virtue of Watson's theorem, share the same phase along the branch cut starting at $s_+ = 4 M_\pi^2$ with the elastic $\pi\pi$ scattering phases up to the respective inelastic thresholds $s_\text{in}^{(l)}$. 
Following Ref.~\cite{Jackson:1968rfn}, we can determine the behavior of the partial-wave amplitudes at the thresholds $s = s_+$, $q^2 = q^2_-$, and $q^2 = q^2_+ \equiv (M_B+ 2 M_\pi)^2$, based on possible kinematic singularities and angular momentum conservation. Again, the polarization and parity of the current play a crucial role. For $f$, $g$, and $\mathcal{F}_2$, we find a simple scaling with $l$:
\begin{align}
    g^{(l)}(q^2,s) &= \left(\sqrt{\lambda_{B\ell}}\beta_\pi\right)^{l-1}\tilde{g}^{(l)}(q^2,s)\,,\nonumber\\
    f^{(l)}(q^2,s) &= \left(\sqrt{\lambda_{B\ell}}\beta_\pi\right)^{l-1}\tilde{f}^{(l)}(q^2,s)\,,\nonumber\\
    \mathcal{F}_2^{(l)}(q^2,s) &= \left(\sqrt{\lambda_{B\ell}}\beta_\pi\right)^l\tilde{\mathcal{F}}_2^{(l)}(q^2,s)\,.\label{eq::ffs_kin_simple}
\end{align}
In the case of $\mathcal{F}_1$ we have to distinguish the S-wave contribution from the others. For $l = 0$ we deal with a $1^+ \rightarrow 0^- 0^+$ transition that can only occur with orbital angular momentum $L = 1$, while the $l = 1$ partial wave is a $1^+ \rightarrow 0^- 1^-$ transition that can proceed with $L = 0$ or $L = 2$. Consequently, we need to include one power of $\sqrt{\lambda_{B\ell}}$ for $l = 0$, but not for $l = 1$~\cite{Jackson:1968rfn}. For higher partial waves, the pattern is the same as for $l = 1$, i.e., transitions with orbital angular momentum $L = l - 1$ and $L = l + 1$ are allowed.
Including kinematic singularities at $q^2 = q^2_\pm$, we obtain 
\begin{align}
    \mathcal{F}_1^{(0)}(q^2,s) &= \tilde{\mathcal{F}}_1^{(0)}(q^2,s)\,,\nonumber\\
    \mathcal{F}_1^{(l)}(q^2,s) &= \frac{1}{\lambda_{B\ell}}\left(\sqrt{\lambda_{B\ell}}\beta_\pi\right)^l\tilde{\mathcal{F}}_1^{(l)}(q^2,s)\,.\label{eq::ffs_kin_simple_2}
\end{align}

In the following, we generalize the derivation of the unitarity bounds for the $B\rightarrow D^{(\ast)}$ form factors by Boyd, Grinstein, and Lebed (BGL)~\cite{Boyd:1995cf,Boyd:1995sq,Boyd:1997kz}. The starting point is the observation that in QCD, the two-point function $\Pi^{(J)}(q^2)_{\mu\nu}$ of currents $J$ obeys once- or twice-subtracted dispersion relations. First, we decompose
\begin{align}
    \Pi^{(J)}(q^2)_{\mu\nu} = P^{(q)}_{\mu\nu}\Pi^{(J)}_T(q^2) + \frac{q_\mu q_\nu}{q^2}\Pi^{(J)}_L(q^2)\,,
\end{align}
where $L$ and $T$ denote the longitudinal and transversal components, respectively. The dispersion relations take the form
\begin{align}
    \chi_L^{(J)}(Q^2) &\equiv \frac{\mathrm{d}\Pi^{(J)}_L}{\mathrm{d}Q^2} = \frac{1}{\pi}\int_0^\infty\mathrm{d}q^2 \frac{\mathrm{Im}\,\Pi^{(J)}_L(q^2)}{(q^2-Q^2)^2}\,, \notag\\
    \chi_T^{(J)}(Q^2) &\equiv \frac{1}{2}\frac{\mathrm{d}^2\Pi^{(J)}_T}{\mathrm{d}(Q^2)^2} = \frac{1}{\pi}\int_0^\infty\mathrm{d}q^2 \frac{\mathrm{Im}\,\Pi^{(J)}_T(q^2)}{(q^2-Q^2)^3}\,,
\end{align}
where the $\chi_{L/T}^{(J)}$ for $b\rightarrow u$ currents can be computed at $Q^2 = 0$ in perturbation theory or on the lattice~\cite{Martinelli:2022tte}. The imaginary parts of $\Pi^{(J)}_{L/T}$ can be expressed through the sum of all possible intermediate hadronic states
\begin{align}
    \mathrm{Im}\,\Pi^{(J)}_{L/T}(q^2 + i\epsilon) =\! \frac{1}{2}\sum_X \int\!\mathrm{dPS} P^{\mu\nu}_{L/T}\left\langle 0 | J_\mu | X \right\rangle\left\langle X | J_\nu | 0 \right\rangle ,
\end{align}
where the projection operators are given by
\begin{align}
    P^{\mu\nu}_{T} = \frac{1}{3}P^{(q),\mu\nu}\quad\text{and}\quad P^{\mu\nu}_{L} = \frac{q^\mu q^\nu}{q^2}\,.
\end{align}

One-particle contributions from poles below the first two-particle threshold to $\mathrm{Im}\,\Pi^{(J)}_{L/T}$ can directly be evaluated, and are given in terms of leptonic decay constants $f_p$ and masses $M_p$:
\begin{align}
    \chi_L^{(J)}(Q^2)\Bigg|_{\text{1-pt}} = \sum_p\frac{M_p^2 f_p^2}{(M_p^2-Q^2)^2}\,,\nonumber\\
    \chi_T^{(J)}(Q^2)\Bigg|_{\text{1-pt}} = \sum_p\frac{M_p^2 f_p^2}{(M_p^2-Q^2)^3}\,.
\end{align}
For $b\rightarrow u$ transitions there are only two subthreshold poles: the $B^\ast$ resonance contributes to the transverse part of the vector current, whereas the $B$ meson contributes to the longitudinal part of the axial current. Similarly, two-particle contributions from $B$ decays to ground-state pseudoscalar mesons will be present for the transverse and longitudinal part of the vector current. However, these will be neglected in the following.
Omitting any intermediate state in the sum leads to an inequality, and thus, an upper bound on the contribution of a given sum of intermediate states to the two-point function.

In our case, $X = B^+\pi^+\pi^-$ and the three-particle phase-space measure can be written as
\begin{align}
    \int\mathrm{dPS}_3 = \int_{s_+}^{s_-}\mathrm{d}s\int_{-1}^1\mathrm{d}\cos\theta_\pi \frac{\sqrt{\lambda_{B\ell}}\beta_\pi}{256\pi^3 q^2}\,,
\end{align}
where $s_- = (M_B-\sqrt{q^2})^2$ depends on $q^2$ and we integrated over the angles that the form factors do not depend on. Inserting our form factor decomposition into the phase-space integrals leads to three contributions: one from $g$ to $\Pi_T^{(V)}$, one from $f$ and $\mathcal{F}_1$ to $\Pi_T^{(A)}$, and one from $\mathcal{F}_2$ to $\Pi_L^{(A)}$. Each of these schematically takes the form
\begin{align}
    \mathrm{Im}\,\Pi(q^2 + i\epsilon)\Bigg|_{B\pi\pi} = \sum_l K_l(q^2,s)|\tilde{F}_l(q^2,s)|^2 \,,
\end{align}
where we integrated over $\cos\theta_\pi$ and collected all numerical and kinematic factors in $K_l(q^2,s)$. Inserting this into the dispersion relations for $Q^2 = 0$ leads to a bound of the form
\begin{align}
    \chi(0) > \frac{1}{\pi}\sum_l\int_{q^2_+}^\infty\mathrm{d}q^2 \int_{s_+}^{s_-}\mathrm{d}s \frac{K_l(q^2,s)}{(q^2)^n}|\tilde{F}_l(q^2,s)|^2\,.
\end{align}
To arrive at a compact BGL-type parameterization of the form factors, we need to disentangle the $s$- and $q^2$-dependence. To this end, we switch the order of integration:
\begin{align}
    \int_{q^2_+}^\infty\mathrm{d}q^2 \int_{s_+}^{s_-}\mathrm{d}s = \int_{s_+}^{\infty}\mathrm{d}s \int_{\tilde{q}^2_+}^\infty\mathrm{d}q^2\,.
\end{align}
The new, $s$-dependent lower $q^2$-integration boundary is given by $\tilde{q}^2_+ = (M_B + \sqrt{s})^2$. We can now write
\begin{align}
    \chi(0) &> \frac{1}{\pi}\sum_l\int_{4 M_\pi^2}^{\infty}\mathrm{d}s \hat{K}_l(s) \nonumber\\ &\otimes\int_{\tilde{q}^2_+}^\infty\mathrm{d}q^2 \frac{\tilde{K}_l(q^2,s)}{(q^2)^n}|\tilde{F}_l(q^2,s)|^2\,,\label{eq::boundsq2}
\end{align}
where we split $K_l(q^2,s)$ into a $q^2$-independent part and a remainder. The $q^2$-integration for fixed $s$ can be treated following the procedure of BGL. This is achieved by mapping the integration domain in $q^2$ onto the unit circle in the variable $z$:
\begin{align}
    z(q^2, q_0^2) = \frac{\sqrt{\hat{q}^2_+ - q^2} - \sqrt{\hat{q}^2_+ - q_0^2}}{\sqrt{\hat{q}^2_+ - q^2} + \sqrt{\hat{q}^2_+ - q_0^2}}\,.
\end{align}
Here, $q^2_0 < \hat{q}^2_+$ determines the value of $q^2$ corresponding to $z = 0$. Note that the branch point of the mapping is the lowest two-particle threshold $\hat{q}^2_+ = (M_{B^{(\ast)}} + M_\pi)^2$, depending on the current under consideration. As a consequence, the integration domain is not the full unit circle, but only an arc with opening angle $\alpha_s = \arg z(\tilde{q}^2_+,q^2_0)$~\cite{Gubernari:2020eft,Blake:2022vfl,Flynn:2023qmi,Gubernari:2023puw}. We thus re-write the integral over $q^2$:
\begin{align}
    &\int_{\tilde{q}^2_+}^\infty\mathrm{d}q^2 \frac{\tilde{K}_l(q^2,s)}{(q^2)^n}|\tilde{F}_l(q^2,s)|^2 \nonumber\\ &= \frac{1}{2 i}\oint\frac{\mathrm{d}z}{z}|\phi_F^{(l)}(z,s) B_F(z) \tilde{F}_l(z,s)|^2\,.
\end{align}
The outer functions $\phi_F^{(l)}(z,s)$ have the same magnitude as the product of the Jacobian of the variable change and kinematic factors on the unit circle, but no zeroes or poles inside of the unit disk, while the Blaschke factors $B_F$ contain subthreshold poles in $q^2$. The integrand is free of kinematic singularities and zeros and thus can be expanded in polynomials orthogonal on the arc of the unit circle. This class of polynomials, characterized by the angle $\alpha_s$, are known as Szeg\H{o} polynomials~\cite{Barry} and their appearance in form factors for semileptonic decays have been first discussed in Ref.~\cite{Gubernari:2020eft}. Expanding the form factors in terms of the Szeg\H{o} polynomials $p_i$ leads to
\begin{align}
\tilde{F}_l(z,s) = \frac{1}{\phi_F^{(l)}(z,s) B_F(z)}\sum_i a^{(F)}_{l,i}(s) p_i(z,\alpha_s)\,,\notag\\
\oint\frac{\mathrm{d}z}{i z}|\phi_F^{(l)}(z,s) B_F(z) \tilde{F}_l(z,s)|^2 = \sum_i |a^{(F)}_{l,i}(s)|^2\,.
\end{align}

Inserting the result into Eq.~\eqref{eq::boundsq2}, we obtain
\begin{align}
    \chi(0) > \frac{1}{2\pi}\sum_{l,i} \int_{s_+}^\infty\mathrm{d}s\, \hat{K}_l(s) |a^{(F)}_{l,i}(s)|^2\,.\label{eq::boundsonly}
\end{align}
In the next step, we need to parameterize the $s$-dependence of the $a^{(F)}_{l,i}(s)$.
Along the branch cut, each of them obeys a unitarity relation of the form
\begin{align}
    \mathrm{Disc}\,a^{(F)}_{l,i}(s) &= \lim_{\epsilon\rightarrow 0}\left(a^{(F)}_{l,i}(s + i\epsilon) - a^{(F)}_{l,i}(s - i\epsilon) \right)\nonumber\\
    &= 2 i a^{(F)}_{l,i}(s) \sin\delta_l\,e^{-i\delta_l}\theta(s-4M_\pi^2)\,,
\end{align}
where the $\delta_l$ are the elastic $\pi\pi$ scattering phases for partial waves with angular momentum $l$. The solution to this equation is given by
\begin{align}
    a^{(F)}_{l,i}(s) &= \Omega_l(s)\tilde{a}^{(F)}_{l,i}(s)\,,\notag\\
    \Omega_l(s) &= \exp\left(\frac{s}{\pi}\int_{s_+}^\infty\mathrm{d}s^\prime\frac{\delta_l(s^\prime)}{s^\prime(s^\prime-s)}\right)\,,
\end{align}
where $\Omega_l(s)$ is the Omnès function~\cite{Omnes:1958hv} and the functions $\tilde{a}^{(F)}_{l,i}(s)$ are real for $4 M_\pi^2 < s < s_\text{in}^{(l)}$.
In Ref.~\cite{Gustafson:2023lrz}, the functions $\tilde{a}^{(F)}_{l,i}(s)$ have been assumed to be approximately $s$-independent.
Here, we aim to expand and derive a general parameterization in $s$, taking into account additional imaginary parts induced above $s_\text{in}^{(l)}$.

The structure of each integral in the sum is exactly of the form as for the pion vector form factor, considered in Ref.~\cite{Caprini:1999ws}, and thus we can resort to the methods introduced there. As we are interested in the region $4 M_\pi^2 \leq s \leq s_\text{in}^{(l)}$, we perform a second conformal mapping,
\begin{align}
    y_l(s,s_0) = \frac{\sqrt{s_\text{in}^{(l)} - s} - \sqrt{s_\text{in}^{(l)} - s_0}}{\sqrt{s_\text{in}^{(l)} - s} + \sqrt{s_\text{in}^{(l)} - s_0}}\,,
\end{align}
where $s_0$ determines the value of $s$ corresponding to $y = 0$.
The variable $y$ is real below the inelastic threshold and lies on the unit circle above.
Splitting the integral in Eq.~\eqref{eq::boundsonly} into two regions, one below the inelastic threshold and one above, we obtain
\begin{align}
    \chi(0) > \frac{1}{2 \pi i}\sum_{l,i}\oint\frac{\mathrm{d}y}{y}|\tilde{\phi}_F^{(l)}(y) \tilde{B}_F^{(l)}(y) \tilde{a}^{(F)}_{l,i}(s)|^2 + \sum_l R_{l}\,.
\end{align}
Here, again, $\tilde{\phi}_F^{(l)}(y)$ are outer functions and $\tilde{B}_F^{(l)}(y)$ possible Blaschke factors. In none of the $\pi\pi$ partial waves subthreshold poles occur, so the Blaschke factors are trivial. However, as we will discuss later, in the $t$-channel P-wave the $B^\ast$ resonance is below the $B\pi$ threshold. We can now express
\begin{align}
    \tilde{a}^{(F)}_{l,i}(s) = \frac{1}{\tilde{\phi}_F^{(l)}(y) \tilde{B}_F^{(l)}(y)}\sum_j c_{l,ij}^{(F)} y^j \,,
\end{align}
where the expansion coefficients $c_{l,ij}^{(F)}$ are real and constrained by
\begin{align}
    \chi(0) > \sum_{l,i,j}|c_{l,ij}^{(F)}|^2 + \sum_{l} R_{l}\,.\label{eq::unitarity1}
\end{align}
The remainders $R_{l}$ are the contribution of the integral of Eq.~\eqref{eq::boundsonly} up to the inelastic threshold and depend non-trivially on the expansion coefficients. While they are positive definite and can be simply evaluated numerically, they are not diagonal in the expansion coefficients $c_{l,ij}^{(F)}$ and mix different powers in the $y$-expansion.

In summary, our partial-wave expanded form factors take the form
\begin{align}
    \tilde{F}^{(l)}(q^2,s) &= \frac{\Omega_l(s)}{\phi_F^{(l)}(z,s) B_F(z) \tilde{\phi}_F^{(l)}(y) \tilde{B}_F^{(l)}(y)}\nonumber\\
    &\otimes\sum_{i,j} c_{l,ij}^{(F)} p_i(z,\alpha_s) y^j\,.
\end{align}
Note that this representation does not benefit from approximate knowledge of the scattering phases above the inelastic threshold nor does it reproduce the correct scaling of imaginary parts stemming from inelastic channels. To ameliorate both issues, we introduce a second form inspired by the Bourrely--Caprini--Lellouch (BCL) parameterization of the $B\rightarrow \pi$ form factors~\cite{Bourrely:2008za}:
\begin{align}
    \tilde{F}^{(l)}(q^2,s) &= \frac{\Omega_l(s)}{(1 - s/M_R^2) \phi_F^{(l)}(z,s) B_F(z)}\nonumber\\
    &\otimes\sum_{i,j} c_{l,ij}^{(F)} p_i(z,\alpha_s) q^{(l)}_j(y)\,.
\end{align}
Here, the Blaschke factors are replaced by explicit pole terms and the outer functions in $y$ are dropped. The polynomials $q^{(l)}_j(y)$ are designed to reproduce the correct scaling at the inelastic threshold, i.e., $\mathrm{Im}\,q^{(l)}_j(y) \propto \sqrt{s_\text{in} - s}^{2l+1}$. While this second parameterization has advantageous analytic properties, the unitarity bound becomes more complicated than in Eq.~\eqref{eq::unitarity1}. In particular, there is no more approximately diagonal structure in the $j$-summation. However, given that the bound in Eq.~\eqref{eq::unitarity1} is non-diagonal in the first place, this does not lead to further complications in practice. A possible future alternative is the inclusion of above-threshold resonances through explicit pole terms, following a first study of the pion vector form factor in Ref.~\cite{Kirk:2024oyl}.

\subsection{Reconstruction theorems}\label{sec::reco}
To include left-hand cuts in $s$ due to $t$- and $u$-channel branch cuts, the discussion of the previous section needs to be extended. The basic derivation follows Refs.~\cite{Kang:2013jaa,Colangelo:2015kha}, while taking the $z$- and $y$-expansion into consideration. Both works follow the Khuri--Treiman (KT) formalism~\cite{Khuri:1960zz}, first introduced to describe $K\rightarrow3\pi$ decays, to take into account two-particle rescattering.

In the KT formalism, decay amplitudes are written as a sum of single-variable amplitudes (SVAs), i.e., amplitudes that depend on either $s$, $t$, or $u$ with prefactors that can depend on the other variables. The prefactors are combinations of phase-space factors and functions of the helicity angles, such as the combinations $(\sqrt{\lambda_{B\ell}}\beta_\pi)^l P_l(\cos\theta_\pi)$, introduced in the previous section. The SVAs themselves have an explicit dependence on the relevant scattering phases, i.e., isoscalar or isovector $\pi\pi$ $l$-wave scattering phases for the $s$-channel and isospin $1/2$ or $3/2$ $B\pi$ $l$-wave scattering phases for the $t$- and $u$-channels.

To obtain SVAs with the correct scattering phases below the first inelastic threshold, we need to decompose them by isospin. To this end, we consider the three $s$-channel $J B \rightarrow \pi \pi$ scattering amplitudes in the physical basis:
\begin{align}
    \mathcal{M}^\mu_{+-} &= \left\langle \pi^+(p_1) \pi^-(p_2) | J^\mu(q) | B^+(p_3) \right\rangle\,,\nonumber\\
    \mathcal{M}^\mu_{00} &= \left\langle \pi^0(p_1) \pi^0(p_2) | J^\mu(q) | B^+(p_3) \right\rangle\,,\nonumber\\
    \mathcal{M}^\mu_{-0} &= \left\langle \pi^-(p_1) \pi^0(p_2) | J^\mu(q) | B^0(p_3) \right\rangle\,.
\end{align}
To obtain their isospin decomposition, we study the three different possible crossings for $2\rightarrow 2$ scattering. In the $s$-channel, we obtain
\begin{align}
    \mathcal{M}^\mu_{J^- B^+ \rightarrow \pi^+ \pi^-} &= \frac{1}{2}\mathcal{M}^{(1),\mu} + \frac{1}{\sqrt{6}}\mathcal{M}^{(0),\mu}\,,\nonumber\\
    \mathcal{M}^\mu_{J^- B^+ \rightarrow \pi^0 \pi^0} &= -\frac{1}{\sqrt{6}}\mathcal{M}^{(0),\mu}\,,\nonumber\\
    \mathcal{M}^\mu_{J^- B^0 \rightarrow \pi^- \pi^0} &= -\frac{1}{\sqrt{2}}\mathcal{M}^{(1),\mu}\,,
\end{align}
where the isovector and isoscalar amplitudes, $\mathcal{M}^{(1),\mu}$ and $\mathcal{M}^{(0),\mu}$, are anti-symmetric and symmetric under exchange of the pions, respectively.
In the $t$- and $u$-channels we obtain
\begin{align}
    \mathcal{M}^\mu_{J^- \pi^+ \rightarrow \pi^+ B^-} &= \frac{1}{3}\mathcal{M}^{(3/2),\mu} + \frac{2}{3}\mathcal{M}^{(1/2),\mu}\,,\nonumber\\
    \mathcal{M}^\mu_{J^- \pi^0 \rightarrow \pi^0 B^-} &= \frac{2}{3}\mathcal{M}^{(3/2),\mu} + \frac{1}{3}\mathcal{M}^{(1/2),\mu}\,,\nonumber\\
    \mathcal{M}^\mu_{J^- \pi^0 \rightarrow \pi^- \bar{B}^0} &= \frac{\sqrt{2}}{3}\mathcal{M}^{(3/2),\mu} - \frac{\sqrt{2}}{3}\mathcal{M}^{(1/2),\mu} \,,
\end{align}
and
\begin{align}
    \mathcal{M}^\mu_{J^- \pi^- \rightarrow \pi^- B^-} &= \mathcal{M}^{(3/2),\mu}\,,\nonumber\\
    \mathcal{M}^\mu_{J^- \pi^0 \rightarrow \pi^0 B^-} &= \frac{2}{3}\mathcal{M}^{(3/2),\mu} + \frac{1}{3}\mathcal{M}^{(1/2),\mu}\,,\nonumber\\
    \mathcal{M}^\mu_{J^- \pi^+ \rightarrow \pi^0 \bar{B}^0} &= \frac{\sqrt{2}}{3}\mathcal{M}^{(3/2),\mu} - \frac{\sqrt{2}}{3}\mathcal{M}^{(1/2),\mu}\,.
\end{align}
For the physical amplitudes we cross back to $J\rightarrow B\pi\pi$ amplitudes and thus obtain
\begin{align}
    \mathcal{M}^\mu_{+-} &= \frac{1}{2}\mathcal{M}^{(1),\mu}(s) + \frac{1}{\sqrt{6}}\mathcal{M}^{(0),\mu}(s)\nonumber\\
    &- \frac{1}{3}\left(2 \mathcal{M}^{(3/2),\mu}(t) + \mathcal{M}^{(1/2),\mu}(t) + (t \leftrightarrow u)\right)\nonumber\\
    &+ \frac{1}{3}\left(\mathcal{M}^{(3/2),\mu}(t) - \mathcal{M}^{(1/2),\mu}(t) - (t \leftrightarrow u)\right)\,,\nonumber\\
    \mathcal{M}^\mu_{00} &= -\frac{1}{\sqrt{6}}\mathcal{M}^{(0),\mu}(s)\nonumber\\
    &+ \frac{1}{3}\left(2 \mathcal{M}^{(3/2),\mu}(t) + \mathcal{M}^{(1/2),\mu}(t) + (t \leftrightarrow u)\right)\,,\nonumber\\
    \mathcal{M}^\mu_{-0} &= \frac{1}{\sqrt{2}}\mathcal{M}^{(1),\mu}(s)\nonumber\\
    &+ \frac{\sqrt{2}}{3}\left(\mathcal{M}^{(3/2),\mu}(t) - \mathcal{M}^{(1/2),\mu}(t) - (t \leftrightarrow u)\right)\,,
\end{align}
where we indicated the Mandelstam variable relevant to the final-state pair. However, the amplitudes themselves at this stage still depend on the other two variables and are not SVAs.

Note that $\mathcal{M}^\mu_{00}$ is completely symmetric under exchange of $p_1$ and $p_2$, whereas $\mathcal{M}^\mu_{-0}$ is completely anti-symmetric.
The remaining amplitude, $\mathcal{M}^\mu_{+-}$ is the most complex due to having mixed symmetry. However, it can be cleanly decomposed into a symmetric and an anti-symmetric part:
\begin{align}
    \mathcal{M}^\mu_{+-} = \frac{\mathcal{M}^\mu_{-0}}{\sqrt{2}} - \mathcal{M}^\mu_{00}\,.
\end{align}
Consequently, going forward we study the symmetric and anti-symmetric amplitudes $\mathcal{M}^\mu_{00}$ and $\mathcal{M}^\mu_{-0}$ separately.

We are now in the position to write the amplitudes with definite isospin in terms of SVAs. The amplitudes can be expressed through the form factors introduced in Sec.~\ref{sec::Setup}, which then can be further written in terms of a single variable. As the tensorial structure multiplying $g$ and $\mathcal{F}_2$ are anti-symmetric and symmetric with respect to exchange of any of the hadron momenta, respectively, their decomposition is straightforward:
\begin{widetext}
\begin{align}
&g^{(-0)}(s,t,u) = \frac{1}{\sqrt{2}}\!\sum_{l\,\text{odd}}\kappa_s^{l-1} P^\prime_l(\cos\theta_\pi)\tilde{g}^{(1)}_{l}(q^2,s) - \frac{\sqrt{2}}{3}\!\sum_l \left(\kappa_t^{l-1} P^\prime_l(\cos\theta_{B1})\left(\tilde{g}^{(1/2)}_{l}(q^2,t)-\tilde{g}^{(3/2)}_{l}(q^2,t)\right) - (t\!\leftrightarrow\!u)\right)\,,\notag\\
    &g^{(00)}(s,t,u) = -\frac{1}{\sqrt{6}}\!\sum_{l\,\text{even}} \! \kappa_s^{l-1} P^\prime_l(\cos\theta_\pi)\tilde{g}^{(0)}_{l}(q^2,s) + \frac{1}{3}\!\sum_l \left(\kappa_t^{l-1} P^\prime_l(\cos\theta_{B1})\left(\tilde{g}^{(1/2)}_{l}(q^2,t)+2\tilde{g}^{(3/2)}_{l}(q^2,t)\right) + (t\!\leftrightarrow\!u)\right) ,\notag\\
    &\mathcal{F}^{(-0)}_2(s,t,u) = \frac{1}{\sqrt{2}}\!\sum_{l\,\text{odd}}\kappa_s^l P_l(\cos\theta_\pi)\tilde{\mathcal{F}}^{(1)}_{2,l}(q^2,s) - \frac{\sqrt{2}}{3}\!\sum_l \left(\kappa_t^l P_l(\cos\theta_{B1})\left(\tilde{\mathcal{F}}^{(1/2)}_{2,l}(q^2,t) -\tilde{\mathcal{F}}^{(3/2)}_{2,l}(q^2,t)\right) + (t\!\leftrightarrow\!u)\right)\,,\notag\\
    &\mathcal{F}^{(00)}_2(s,t,u) = -\frac{1}{\sqrt{6}}\!\sum_{l\,\text{even}}\kappa_s^l P_l(\cos\theta_\pi)\tilde{\mathcal{F}}^{(0)}_{2,l}(q^2,s) + \frac{1}{3}\!\sum_l \left(\kappa_t^l P_l(\cos\theta_{B1})\left(\tilde{\mathcal{F}}^{(1/2)}_{2,l}(q^2,t) +2\tilde{\mathcal{F}}^{(3/2)}_{2,l}(q^2,t)\right)- (t\!\leftrightarrow\!u)\right)\,,
\end{align}
\end{widetext}
where we separated the form factors for the $\pi^-\pi^0$ and $\pi^0\pi^0$ channels and
\begin{align}
    &\cos\theta_{B1} = \frac{t(s - u) - (M_B^2 - M_\pi^2)(q^2 - M_\pi^2)}{\kappa_t}\,,\notag\\
    &\cos\theta_{B2} = \frac{u(s - t) - (M_B^2 - M_\pi^2)(q^2 - M_\pi^2)}{\kappa_u}\,,\notag\\
    &\kappa_s = \sqrt{\lambda_{B\ell}}\beta_\pi\,,\quad\kappa_t = \frac{\sqrt{\lambda_{B1}\lambda_{2\ell}}}{t}\,,\quad\kappa_u = \frac{\sqrt{\lambda_{B2}\lambda_{1\ell}}}{u}\,,\notag\\
    &\lambda_{B1} = \lambda(t,M_B^2,M_\pi^2)\,,\quad\lambda_{B2} = \lambda(u,M_B^2,M_\pi^2)\,,\nonumber\\
    &\lambda_{2\ell} = \lambda(t,q^2,M_\pi^2)\,,\quad\lambda_{1\ell} = \lambda(u,q^2,M_\pi^2)\,.
\end{align}
The $u$-channel crossings are simply obtained by the replacements $t\leftrightarrow u$ and $1 \leftrightarrow 2$. The summation over even $l$ for the function $g^{(00)}$ starts at $l = 2$ as $P_0^\prime(x) = 0$.

For the two transversal form factors of the axial current, the situation is complicated by the non-trivial dependence of the tensor structures in the decomposition on two out of the three momenta. While the decomposition proposed in Sec.~\ref{sec::Setup} is advantageous for the physical decay rate, it is less suited for deriving SVAs with clear threshold behavior and symmetry properties. To resolve this issue, we include the tensor structures in the derivation:
\begin{align}
    &P^{(q)}_{\mu\nu}p_{12}^\nu \mathcal{F}_1(s,t,u) + T_{\mu}^{(12)}f(s,t,u) =\nonumber\\
    &P^{(q)}_{\mu\nu}\left(p_{12}^\nu \mathcal{F}^{(s)}_1(s,t,u) + p_{13}^\nu \mathcal{F}^{(t)}_1(s,t,u) + p_{23}^\nu \mathcal{F}^{(u)}_1(s,t,u) \right)\nonumber\\
    &+\!T_{\mu}^{(12)} f^{(s)}(s,t,u)\!+\!T_{\mu}^{(13)} f^{(t)}(s,t,u)\!+\!T_{\mu}^{(23)} f^{(u)}(s,t,u)\,.
\end{align}

The $s$-channel contributions are now given by
\begin{align}
    \mathcal{F}_1^{(-0),(s)}(s,t,u) &= \frac{1}{\sqrt{2}\lambda_{B\ell}}\sum_{l\,\text{odd}}\kappa_s^l P_l(\cos\theta_\pi)\tilde{\mathcal{F}}_{1,l}^{(1)}(q^2,s)\,,\notag\\
    \mathcal{F}_1^{(00),(s)}(s,t,u) &= -\frac{1}{\sqrt{6}}\tilde{\mathcal{F}}_{1,0}^{(0)}(q^2,s)\notag\\&-\frac{1}{\sqrt{6}\lambda_{B\ell}}\sum_{l\,\text{even}}\kappa_s^l P_l(\cos\theta_\pi)\tilde{\mathcal{F}}_{1,l}^{(0)}(q^2,s)\,,\notag\\
    f^{(-0),(s)}(s,t,u) &= \frac{1}{\sqrt{2}}\sum_{l\,\text{odd}}\kappa_s^{l-1} P^\prime_l(\cos\theta_\pi)\tilde{f}_{l}^{(1)}(q^2,s)\,,\notag\\
    f^{(00),(s)}(s,t,u) &= -\frac{1}{\sqrt{6}}\sum_{l\,\text{even}}\kappa_s^{l-1} P^\prime_l(\cos\theta_\pi)\tilde{f}_{l}^{(0)}(q^2,s)\,,
\end{align}
while the $t$-channel contributions take the form
\begin{widetext}
\begin{align}
    \mathcal{F}_1^{(-0),(t)}(s,t,u) &= -\frac{\sqrt{2}}{3}\left(\tilde{\mathcal{F}}_{1,0}^{(1/2)}(q^2,t)-\tilde{\mathcal{F}}_{1,0}^{(3/2)}(q^2,t)\right)-\frac{\sqrt{2}}{3\lambda_{2\ell}}\sum_{l \geq 1}\kappa_t^l P_l(\cos\theta_{B1})\left(\tilde{\mathcal{F}}_{1,l}^{(1/2)}(q^2,t)-\tilde{\mathcal{F}}_{1,l}^{(3/2)}(q^2,t)\right)\,,\notag\\
    \mathcal{F}_1^{(00),(t)}(s,t,u) &= \frac{1}{3}\left(\tilde{\mathcal{F}}_{1,0}^{(1/2)}(q^2,t)+2\tilde{\mathcal{F}}_{1,0}^{(3/2)}(q^2,t)\right)+\frac{1}{3\lambda_{2\ell}}\sum_{l \geq 1}\kappa_t^l P_l(\cos\theta_{B1})\left(\tilde{\mathcal{F}}_{1,l}^{(1/2)}(q^2,t)+2\tilde{\mathcal{F}}_{1,l}^{(3/2)}(q^2,t)\right)\,,\notag\\
    \mathcal{F}_1^{(-0),(t)}(s,t,u) &= -\frac{\sqrt{2}}{3}\left(\tilde{\mathcal{F}}_{1,0}^{(1/2)}(q^2,t)-\tilde{\mathcal{F}}_{1,0}^{(3/2)}(q^2,t)\right)-\frac{\sqrt{2}}{3\lambda_{2\ell}}\sum_{l \geq 1}\kappa_t^l P_l(\cos\theta_{B1})\left(\tilde{\mathcal{F}}_{1,l}^{(1/2)}(q^2,t)-\tilde{\mathcal{F}}_{1,l}^{(3/2)}(q^2,t)\right)\,,\notag\\
    \mathcal{F}_1^{(00),(t)}(s,t,u) &= \frac{1}{3}\left(\tilde{\mathcal{F}}_{1,0}^{(1/2)}(q^2,t)+2\tilde{\mathcal{F}}_{1,0}^{(3/2)}(q^2,t)\right)+\frac{1}{3\lambda_{2\ell}}\sum_{l \geq 1}\kappa_t^l P_l(\cos\theta_{B1})\left(\tilde{\mathcal{F}}_{1,l}^{(1/2)}(q^2,t)+2\tilde{\mathcal{F}}_{1,l}^{(3/2)}(q^2,t)\right)\,,\notag\\
    f^{(-0),(t)}(s,t,u) &= -\frac{\sqrt{2}}{3}\sum_{l}\kappa_t^{l-1} P^\prime_l(\cos\theta_{B1})\left(\tilde{f}_{l}^{(1/2)}(q^2,t)-\tilde{f}_{l}^{(3/2)}(q^2,t)\right)\,,\notag\\
    f^{(00),(t)}(s,t,u) &= \frac{1}{3}\sum_{l}\kappa_t^{l-1} P^\prime_l(\cos\theta_{B1})\left(\tilde{f}_{l}^{(1/2)}(q^2,t)+2\tilde{f}_{l}^{(3/2)}(q^2,t)\right)\,.
\end{align}
\end{widetext}
For all functions we separated the S-wave contribution and thus all sums over even $l$ start at $l = 2$.
The $u$-channel crossings are related to the $t$-channel ones through
\begin{align}
    \mathcal{F}_1^{(-0),(u)}(s,t,u) &= -\mathcal{F}_1^{(-0),(t)}(s,u,t)\,,\nonumber\\
    \mathcal{F}_1^{(00),(u)}(s,t,u) &= \mathcal{F}_1^{(00),(t)}(s,u,t)\,,\nonumber\\
    f^{(-0),(u)}(s,t,u) &= -f^{(-0),(t)}(s,u,t)\,,\nonumber\\
    f^{(00),(u)}(s,t,u) &= f^{(00),(t)}(s,u,t)\,.
\end{align}

\subsection{Unitarity bounds and parameterization}
With these expressions at hand, we can now compute the contributions to the imaginary parts of the two-point functions $\Pi_{L/T}^{(J)}$. All three charge configurations contribute to the bounds and we can write
\begin{align}
    \mathrm{Im}\,\Pi_{L/T}^{(J)} &= \frac{1}{2}\int\mathrm{dPS}_3\,P_{L/T}^{\mu\nu}\Big(\mathcal{M}_{+-,\mu}\mathcal{M}^\ast_{+-,\nu}\nonumber\\&+\mathcal{M}_{00,\mu}\mathcal{M}^\ast_{00,\nu}+\mathcal{M}_{-0,\mu}\mathcal{M}^\ast_{-0,\nu}\Big)\nonumber\\
    &= \frac{1}{2}\int\mathrm{dPS}_3\,P_{L/T}^{\mu\nu}\Big(2\mathcal{M}_{00,\mu}\mathcal{M}^\ast_{00,\nu}\nonumber\\
    &+\frac{3}{2}\mathcal{M}_{-0,\mu}\mathcal{M}^\ast_{-0,\nu} - \sqrt{2}\mathrm{Re}(\mathcal{M}_{-0,\mu}\mathcal{M}^\ast_{00,\nu})\Big)\,.
\end{align}
Note that the last term drops out after angular integration, as the $-0$ and $00$ amplitudes are anti-symmetric and symmetric under exchange of the pions, respectively.

The two other terms contain contributions diagonal in the Mandelstam variables, but also off-diagonal interference terms.
These interference terms only constitute small perturbations on top of the dominant resonant contributions in the diagonal terms and, consequently, we neglect them in the derivation of a suitable form factor parameterization and take them into account through a modification of the unitarity bounds.
Focusing on the diagonal terms and writing the $u$-channel contributions as $t$-channel integrals we obtain
\begin{widetext}
\begin{align}
    \mathrm{Im}\,\Pi_T^{(V)}(q^2 + i\epsilon)&\Bigg|_\text{diag} = \frac{1}{12288\pi^3 q^2} \sum_l \frac{l(l+1)}{2l+1}\Bigg[\int_{s_+}^{s_-}\mathrm{d}s\,s \kappa_s^{2l + 1}\left(\frac{1}{4}|\tilde{g}_l^{(0)}|^2 + \frac{3}{4}|\tilde{g}_l^{(1)}|^2\right) \nonumber\\+ &\int_{t_+}^{t_-}\mathrm{d}t\,t \kappa_t^{2l + 1}\left(|\tilde{g}_l^{(1/2)}|^2  + 2|\tilde{g}_l^{(3/2)}|^2\right)\Bigg]\,,\notag\\
    \mathrm{Im}\,\Pi_L^{(A)}(q^2 + i\epsilon)&\Bigg|_\text{diag} = \frac{1}{256\pi^3 q^4} \sum_l \frac{1}{2l+1}\Bigg[\int_{s_+}^{s_-}\mathrm{d}s\,\kappa_s^{2l + 1}\left(\frac{1}{4}|\tilde{\mathcal{F}}_{2,l}^{(0)}|^2 + \frac{3}{4}|\tilde{\mathcal{F}}_{2,l}^{(1)}|^2\right) \nonumber\\+ &\int_{t_+}^{t_-}\mathrm{d}t\,\kappa_t^{2l + 1}\left(|\tilde{\mathcal{F}}_{2,l}^{(1/2)}|^2  + 2|\tilde{\mathcal{F}}_{2,l}^{(3/2)}|^2\right)\Bigg]\,,\notag\\
    \mathrm{Im}\,\Pi_T^{(A)}(q^2 + i\epsilon)&\Bigg|_\text{diag} = \frac{1}{3072\pi^3 q^4} \Bigg\{\int_{s_+}^{s_-}\mathrm{d}s\, \frac{\kappa_s}{4}\lambda_{B\ell}|\tilde{\mathcal{F}}_{1,0}^{(0)}|^2 + \sum_{l>0} \frac{1}{2l+1}\int_{s_+}^{s_-}\mathrm{d}s\,\frac{\kappa_s^{2l + 1}}{\lambda_{B\ell}}\left(\frac{1}{4}|\tilde{\mathcal{F}}_{1,l}^{(0)}|^2 + \frac{3}{4}|\tilde{\mathcal{F}}_{1,l}^{(1)}|^2\right)\nonumber\\ + &\int_{t_+}^{t_-}\mathrm{d}t\,\kappa_t\lambda_{2\ell}\left(|\tilde{\mathcal{F}}_{1,0}^{(1/2)}|^2  + 2|\tilde{\mathcal{F}}_{1,0}^{(3/2)}|^2\right)+\sum_{l>0}\frac{1}{2l+1}\int_{t_+}^{t_-}\mathrm{d}t\, \frac{\kappa_t^{2l+1}}{\lambda_{2\ell}}\left(|\tilde{\mathcal{F}}_{1,l}^{(1/2)}|^2  + 2|\tilde{\mathcal{F}}_{1,l}^{(3/2)}|^2\right)\nonumber\\
    + &4 q^2\sum_l \frac{l(l+1)}{2l+1}\Bigg[\int_{s_+}^{s_-}\mathrm{d}s\, \frac{s\kappa_s^{2l + 1}}{\lambda_{B\ell}}\left(\frac{1}{4}|\tilde{f}_{l}^{(0)}|^2 + \frac{3}{4}|\tilde{f}_{l}^{(1)}|^2\right)+\int_{t_+}^{t_-}\mathrm{d}t\, \frac{t\kappa_t^{2l + 1}}{\lambda_{2\ell}}\left(|\tilde{f}_{l}^{(1/2)}|^2 + 2|\tilde{f}_{l}^{(3/2)}|^2\right)\Bigg]\Bigg\}\,.
    \label{eq::unitarity2}
\end{align}
\end{widetext}
The $t$- and $u$-channel integration boundaries are given by
\begin{align}
&t_+ = u_+ = (M_B+M_\pi)^2\, ,\notag\\
&t_- = u_- = (M_\pi - \sqrt{q^2})^2\,.
\end{align}

In the next step, we derive a parameterization for each of the form factors. To simplify the discussion, we focus on $\mathrm{Im}\,\Pi_L^{(A)}$, but the other three form factors follow in a similar manner.
Inserting the imaginary part of the two-point function into the dispersion relation yields
\begin{widetext}
\begin{align}
    \chi^{A}_L(0) \geq &\sum_l\frac{2N_l}{\pi}\int_{q^2_+}^\infty\frac{\mathrm{d}q^2}{q^8}\Bigg[\int_{s_+}^{s_-}\mathrm{d}s\, \kappa_s^{2l + 1}\left(\frac{1}{4}|\tilde{\mathcal{F}}_{2,l}^{(0)}|^2 + \frac{3}{4}|\tilde{\mathcal{F}}_{2,l}^{(1)}|^2\right) +\int_{t_+}^{t_-}\mathrm{d}t\, \kappa_t^{2l + 1}\left(|\tilde{\mathcal{F}}_{2,l}^{(1/2)}|^2 + 2|\tilde{\mathcal{F}}_{2,l}^{(3/2)}|^2\right)\Bigg]\nonumber\\
    = &\sum_l\frac{2N_l}{\pi}\Bigg[\int_{s_+}^{\infty}\mathrm{d}s\,\beta_\pi^{2l+1}\int_{\tilde{q}^2_{+,s}}^\infty\mathrm{d}q^2 \frac{\sqrt{\lambda_{B\ell}}^{2l+1}}{q^8}\left(\frac{1}{4}|\tilde{\mathcal{F}}_{2,l}^{(0)}|^2 + \frac{3}{4}|\tilde{\mathcal{F}}_{2,l}^{(1)}|^2\right)\nonumber\\
    &\qquad+\int_{t_+}^{\infty}\mathrm{d}t\, \beta_B^{2l+1}\int_{\tilde{q}^2_{+,t}}^\infty\mathrm{d}q^2 \frac{\sqrt{\lambda_{2\ell}}^{2l+1}}{q^8}\left(|\tilde{\mathcal{F}}_{2,l}^{(1/2)}|^2 + 2|\tilde{\mathcal{F}}_{2,l}^{(3/2)}|^2\right)\Bigg]\,,
\end{align}
\end{widetext}
where $N_l^{-1} = 512\pi^3(2l+1)$, $\beta_B = \sqrt{\lambda_{B1}}/t$, $\tilde{q}^2_{+,s} = (M_B + \sqrt{s})^2$, and $\tilde{q}^2_{+,t} = (M_\pi + \sqrt{t})^2$.
The $q^2$-integration can be approached with standard techniques. We express $q^2$ through
\begin{align}
    z(q^2,q^2_0) = \frac{\sqrt{\hat{q}^2_+ - q^2} - \sqrt{\hat{q}^2_+ - q_0^2}}{\sqrt{\hat{q}^2_+ - q^2} + \sqrt{\hat{q}^2_+ - q_0^2}}
\end{align}
and replace the kinematic factors depending on $q^2$ through outer functions:
\begin{align}
    q^2 - x_i&\rightarrow \phi_i(q^2) = \frac{x_i - q^2}{z(q^2,x_i)}\,.
\end{align}
In contrast to the two-particle case, the $x_i$ depend not only on particle masses, but also on $s$ or $t$.
The full set is given by all values of $q^2$ where kinematic factors can vanish:
\begin{align}
    x_0 &= 0\,,\nonumber\\
    x_{1/2} &= (M_B \pm \sqrt{s})^2\,,\nonumber\\
    x_{3/4} &= (M_\pi \pm \sqrt{t})^2\,.
\end{align}
Some of the $x_i$ might be larger than $\hat{q}^2_+$, depending on the kinematic region under consideration. In that case, the denominator of the corresponding outer function reduces to unity.
The outer functions are given by
\begin{align}
    |\bar{\phi}_{\mathcal{F}_2,l}^{(0)}(q^2,s)|^2 &= \frac{\eta^{(0)}N_l}{\chi^{A}_L(0)}\left|\frac{\mathrm{d}q^2}{\mathrm{d}z}\right|\frac{\sqrt{\phi_1\phi_2}^{2l+1}}{\phi_0^4}\,,\nonumber\\
    |\bar{\phi}_{\mathcal{F}_2,l}^{(1)}(q^2,s)|^2 &= \frac{\eta^{(1)}N_l}{\chi^{A}_L(0)}\left|\frac{\mathrm{d}q^2}{\mathrm{d}z}\right|\frac{\sqrt{\phi_1\phi_2}^{2l+1}}{\phi_0^4}\,,\nonumber\\
    |\bar{\phi}_{\mathcal{F}_2,l}^{(1/2)}(q^2,t)|^2 &= \frac{\eta^{(1/2)}N_l}{\chi^{A}_L(0)}\left|\frac{\mathrm{d}q^2}{\mathrm{d}z}\right|\frac{\sqrt{\phi_3\phi_4}^{2l+1}}{\phi_0^4}\,,\nonumber\\
    |\bar{\phi}_{\mathcal{F}_2,l}^{(3/2)}(q^2,t)|^2 &= \frac{\eta^{(3/2)}N_l}{\chi^{A}_L(0)}\left|\frac{\mathrm{d}q^2}{\mathrm{d}z}\right|\frac{\sqrt{\phi_3\phi_4}^{2l+1}}{\phi_0^4}\,,
\end{align}
where the isospin factors are given by
\begin{align}
    \eta^{(0)} &= \frac{1}{4}\,, & \eta^{(1)} &= \frac{3}{4}\,,\nonumber\\\eta^{(1/2)} &= 1\,, & \eta^{(3/2)} &= 2\,.
\end{align}
In addition, we need to introduce a Blaschke factor $B_{\mathcal{F}_2}$ to cancel the subthreshold pole at $q^2 = M_B^2$:
\begin{align}
    B_{\mathcal{F}_2}(q^2,q^2_0) = \frac{z(q^2,q^2_0) - z(M_B^2,q^2_0)}{1-z(q^2,q^2_0)z(M_B^2,q^2_0)}\,.
\end{align}
The bound now takes the form
\begin{widetext}
\begin{align}
    1 \geq \frac{1}{\pi}\sum_l\Bigg[&\int_{s_+}^{\infty}\mathrm{d}s\,\beta_\pi^{2l+1}\oint \frac{\mathrm{d}z}{z}\left(|\bar{\phi}_{\mathcal{F}_2,l}^{(0)}B_{\mathcal{F}_2}\tilde{\mathcal{F}}_{2,l}^{(0)}|^2 + |\bar{\phi}_{\mathcal{F}_2,l}^{(1)}B_{\mathcal{F}_2}\tilde{\mathcal{F}}_{2,l}^{(1)}|^2\right)\theta(\alpha_s - |\arg(z)|)\nonumber\\
    &+\int_{t_+}^{\infty}\mathrm{d}t\,\beta_B^{2l+1}\oint \frac{\mathrm{d}z}{z}\Big(|\bar{\phi}_{\mathcal{F}_2,l}^{(1/2)}B_{\mathcal{F}_2}\tilde{\mathcal{F}}_{2,l}^{(1/2)}|^2  +  |\bar{\phi}_{\mathcal{F}_2,l}^{(3/2)}B_{\mathcal{F}_2}\tilde{\mathcal{F}}_{2,l}^{(3/2)}|^2\Big)\theta(\alpha_t - |\arg(z)|)\Bigg]\,.
\end{align}
\end{widetext}
Here $\alpha_s = |\arg(z(\tilde{q}^2_{+,s}))|$ and $\alpha_t = |\arg(z(\tilde{q}^2_{+,t}))|$.
The $z$ integration can now be performed by expressing the form factors through
\begin{align}
    \tilde{\mathcal{F}}_{2,l}^{(I)}(q^2,x) = \frac{1}{B_{\mathcal{F}_2}(q^2)\bar{\phi}_{\mathcal{F}_2,l}^{(I)}(q^2,x)}\sum_i d^{(I)}_{l,i}(x) p_i(q^2,\alpha_x)\,,
\end{align}
where the $p_i$ are the Szeg\H{o} polynomials. Note that for semileptonic decays
\begin{align}
   t = \frac{1}{2}\left(M_B^2 + 2 M_\pi^2 + q^2 -s +\kappa_{12}\cos\theta_\pi\right) \leq M_B^2
\end{align}
and thus $\alpha_t$ is not necessarily well defined, as $z(\tilde{q}^2_{+,t})$ takes on real values. To analytically continue the Szeg\H{o} polynomials in this scenario we first observe that for $\tilde{q}^2_{+,t} = \hat{q}^2_+$ the integration covers the full unit circle and the Szeg\H{o} polynomials simply reduce to monomials in $z$. Decreasing $\tilde{q}^2_{+,t}$ further transforms the integration contour to a one-sided keyhole contour and thus the analytic continuation of the Szeg\H{o} polynomials can be obtained by numerical orthogonalization on the contour and matching to monomials for $\tilde{q}^2_{+,t} = \hat{q}^2_+$.

Inserting the expansion into the bound and performing the integral yields
\begin{align}
    1 \geq \frac{1}{\pi}\sum_{l,i}\Bigg[\int_{s_+}^{\infty}\mathrm{d}s\,\beta_\pi^{2l+1} \left(|d_{l,i}^{(0)}|^2 + |d_{l,i}^{(1)}|^2\right)\nonumber\\ +\int_{t_+}^{\infty}\mathrm{d}t\,\beta_B^{2l+1}\left(|d_{l,i}^{(1/2)}|^2 + |d_{l,i}^{(3/2)}|^2\right)\Bigg]\,.\label{eq::bound1}
\end{align}

As the $B\pi$ scattering phases are mostly unknown, we map the whole region of $t > t_+$ onto the unit circle:
\begin{align}
    y_t(t,t_0) = \frac{\sqrt{t_+ - t} - \sqrt{t_+- t_0}}{\sqrt{t_+ - t} + \sqrt{t_+- t_0}}\,.
\end{align}
The isospin-$1/2$ P-wave has one subthreshold pole, the $B^\ast$ resonance, that needs to be taken into account through a Blaschke factor:
\begin{align}
    \tilde{B}^{(1/2)}_{\mathcal{F}_2,1}(t,t_0) = \frac{y_t(t,t_0) - y_t(M_{B^\ast}^2,t_0)}{1-y_t(t,t_0)y_t(M_{B^\ast}^2,t_0)}\,.
\end{align}
Analogously to the $q^2$-dependence we introduce outer functions, which in this case only depend on the particle masses:
\begin{align}
    t - \tilde{x}_i \rightarrow \tilde{\phi}_i &= \frac{\tilde{x}_i - t}{y_t(t,\tilde{x}_i)}\,,\nonumber\\
    \tilde{x}_0 = 0\,,\quad \tilde{x}_{1} &= (M_B - M_\pi)^2\,.
\end{align}
Consequently the full $t$-dependent outer functions are
\begin{align}
    |\hat{\phi}^{(I)}_{\mathcal{F}_2,l,i}(t)|^2 = \left|\frac{\mathrm{d}t}{\mathrm{d}y_t}\right|\left(\frac{\sqrt{t - t_+}\sqrt{\tilde{\phi}_1}}{\tilde{\phi}_0}\right)^{2l+1}\,.
\end{align}
Finally, we can write
\begin{align}
    d_{l,i}^{(1/2)}(t) &= \frac{1}{\hat{\phi}^{(1/2)}_{\mathcal{F}_2,l,i}(t)\tilde{B}^{(1/2)}_{\mathcal{F}_2,l}}\sum_j d^{(1/2)}_{l,ij} y_t^j\,,\notag\\
    d_{l,i}^{(3/2)}(t) &= \frac{1}{\hat{\phi}^{(3/2)}_{\mathcal{F}_2,l,i}(t)}\sum_j d^{(3/2)}_{l,ij} y_t^j\,.
\end{align}

The $s$-dependent form factors, on the other hand, can be decomposed into two parts, the Omnès function and a piece containing crossed-channel contributions and inelasticities:
\begin{align}
    d_{l,i}^{(I)}(s) = \Omega_l(s)\bar{d}^{(I)}_{l,i}(s)\,.
\end{align}
The function $\bar{d}^{(I)}_{l,i}(s)$ is induced by rescattering and real below threshold, but acquires an imaginary part above. It can be related to the $t$- and $u$-channel contributions through
\begin{align}
    \bar{d}^{(I)}_{l,i}(s) &= \tilde{d}^{(I)}_{l,i}(s) + X_{l,i}^{(I)}(s)\,,\nonumber\\
    X_{l,i}^{(I)}(s) &= \frac{s}{\pi}\int_{s_+}^\infty\mathrm{d}s^\prime\frac{\sin\delta_l(s^\prime)\hat{d}_{l,i}^{(I)}(s^\prime)}{|\Omega_l(s^\prime)|s^\prime(s^\prime-s)}\,,
\end{align}
where
\begin{align}
    \hat{d}_{l,i}^{(I)}(s) &+ d^{(I)}_{l,i}(s) = \oint\frac{\mathrm{d}z}{z} p_i^\ast(z,\alpha_s)\nonumber\\ &\otimes\frac{2l+1}{2\kappa_s^l}\int_{-1}^1\mathrm{d}\cos\theta_\pi\, P_l(\cos\theta_\pi)\mathcal{F}_2(s,t,u)\,.
\end{align}
Note that the $\hat{d}_{l,i}^{(I)}(s)$ represent the contributions of the $t$- and $u$-channel SVAs to the $s$-channel partial wave projections.

Consequently, the only undetermined piece is $\tilde{d}^{(I)}_{l,i}(s)$ that we will express through
\begin{align}
    y_{s}(s,s_0) = \frac{\sqrt{s^{(l)}_\text{in} - s} - \sqrt{s^{(l)}_\text{in} - s_0}}{\sqrt{s^{(l)}_\text{in} - s} + \sqrt{s^{(l)}_\text{in} - s_0}}\,,
\end{align}
where the position of the first inelastic threshold, $s_\text{in}$, can differ between partial waves.
The outer functions are given by
\begin{align}
    |\hat{\phi}^{(I)}_{\mathcal{F}_2,l,i}|^2 = \left|\frac{\mathrm{d}s}{\mathrm{d}y_s}\right|
    \left(\frac{\beta_\pi \sqrt{y_s(s,0)}}{\sqrt{y_s(s,s_+)}}\right)^{2l+1}
    |\phi^{(I)}_{\Omega,l}|^2\,,
\end{align}
where $\phi^{(I)}_{\Omega,l}$ is the outer function corresponding to the Omnès function and is given by
\begin{align}
    \phi^{(I)}_{\Omega,l}(y_s) = e^{i\varphi}\exp\left(\frac{1}{2\pi i}\oint\frac{\mathrm{d}y}{y}\frac{y+y_s}{y-y_s}\ln\left(|\Omega_l(y)|\right)\right)\,.
\end{align}
Here, the phase $\varphi$ is arbitrary and by definition $|\tilde{\phi}^{(I)}_{\Omega,l}(y_s)| = \Omega_l(y_s)$ for $|y_s| = 1$. Consequently, the outer function cancels the Omnès factor above the inelastic threshold up to a relative phase.

Combining all pieces, we obtain
\begin{align}
    \tilde{d}^{(I)}_{l,i}(s) = \frac{1}{\hat{\phi}^{(I)}_{\mathcal{F}_2,l,i}(s)}\sum_j d^{(I)}_{l,ij} y_s^j\,.
\end{align}
Therefore, the bound in Eq.~\eqref{eq::bound1} takes the form
\begin{align}
    1 &\geq \sum_l R_l + \sum_{l,ij}\Big(\hat{R}_{l,ij} + \tilde{R}_{l,ij} \Big)\nonumber\\ &+ \sum_{l,i,j}\Big(|d^{(0)}_{l,ij}|^2 + |d^{(1)}_{l,ij}|^2 + |d^{(1/2)}_{l,ij}|^2 + |d^{(3/2)}_{l,ij}|^2\Big)\,.
\end{align}
The remainders are given by
\begin{align}
    R_l &= \frac{1}{\pi}\sum_i \int_{s_+}^{s_\text{in}}\mathrm{d}s\,\beta_\pi^{2l+1} \left(|d_{l,i}^{(0)}|^2 + |d_{l,i}^{(1)}|^2\right)\,,\nonumber\\
    \hat{R}_{l,ij} &= \frac{1}{\pi} \int_{s_+}^{\infty}\mathrm{d}s\,\beta_\pi^{2l+1}\Big(X_{l,i}^{(0)}X_{l,j}^{(0),\ast} + X_{l,i}^{(1)}X_{l,j}^{(1),\ast}\Big)\,,\nonumber\\
    \tilde{R}_{l,ij} &= \frac{2}{\pi} \int_{s_+}^{\infty}\mathrm{d}s\,\beta_\pi^{2l+1}\mathrm{Re}\Big(\tilde{d}^{(0),\ast}_{l,i} X_{l,j}^{(0)}+\tilde{d}^{(1),\ast}_{l,i} X_{l,j}^{(1)}\Big)\,.
\end{align}
Finally, the terms off-diagonal in the $s$-, $t$-, and $u$-channel contributions must be added and numerically integrated.

The derivation of the bounds for $g$, $\mathcal{F}_1$, and $f$ proceeds in the same manner, the differences in kinematic and combinatorial factors lead to slight differences in the outer functions. In summary, we can write all outer functions in the form
\begin{widetext}
\begin{align}
    |\phi^{(I)}_{F,l,i}(q^2,x)|^2 &= |\tilde{\phi}^{(I)}_{F,l}(q^2,x)\hat{\phi}^{(I)}_{F,l,i}(x)|^2 = n_l \frac{\eta^{(I)}N_l}{\chi(0)}\left|\frac{\mathrm{d}q^2}{\mathrm{d}z}\right|\left|\frac{\mathrm{d}x}{\mathrm{d}y_x}\right||\phi_{z,l}(q^2,x)|^2 |\phi_{x,l}(x)|^2 |\phi^{(I)}_{\Omega,l}(x)|^2\,,\notag\\
    |\phi_{z,l}(q^2,s)|^2 &=\left(\frac{-z(q^2,0)}{q^2}\right)^a \left(\sqrt{\frac{(q^2 - (M_B-\sqrt{s})^2)(q^2 - (M_B+\sqrt{s})^2)}{z(q^2,(M_B-\sqrt{s})^2)z(q^2,(M_B+\sqrt{s})^2)}}\right)^{2l + 1 - b}\,,\notag\\
    |\phi_{z,l}(q^2,t)|^2 &=\left(\frac{-z(q^2,0)}{q^2}\right)^a \left(\sqrt{\frac{(q^2 - (M_\pi-\sqrt{t})^2)(q^2 - (M_\pi+\sqrt{t})^2)}{z(q^2,(M_\pi-\sqrt{t})^2)z(q^2,(M_\pi+\sqrt{t})^2)}}\right)^{2l + 1 - b}\,,\notag\\
    |\phi_{s,l}(s)|^2 &= \left(\beta_\pi\frac{\sqrt{y_s(s,0)}}{\sqrt{y_s(s,s_+)}}\right)^{2l+1}\left(\frac{-s}{y_s(s,0)}\right)^c\,,\notag\\
    |\phi_{t,l}(t)|^2 &= \left(\beta_B\frac{y_t(t,0)}{\sqrt{y_t(t,(M_B-M_\pi)^2)}}\right)^{2l+1}\left(\frac{-t}{y_t(t,0)}\right)^c\,.
\end{align}
\end{widetext}
The parameters $n$, $a$, $b$, and $c$ are given in Table~\ref{tbl::outer}.

\begin{table}
\caption{Parameters of the outer functions for a given form factor $F$.}
\label{tbl::outer}
\renewcommand{\arraystretch}{1.5}
\begin{tabular}{l|ccccc}
\hline\hline
     $F$ & $\chi(0)$ & $n_l$ & $a$ & $b$ & $c$\\
     \hline
     $g$ & $\chi_T^{(V)}(0)$ & $\frac{1}{48}l(l+1)$ & 4 & 0 & 1\\
     $f$ & $\chi_T^{(A)}(0)$ & $\frac{1}{3}l(l+1)$ & 4 & 2 & 1\\
     $\mathcal{F}_1$ & $\chi_T^{(A)}(0)$ & $\frac{1}{12}$ & 5 & 2 & 0\\
     $\mathcal{F}_2$ & $\chi_L^{(A)}(0)$ & 1 & 4 & 0 & 0\\
     \hline\hline
\end{tabular}
\end{table}

An alternative parameterization can be derived for $\bar{d}_{l,i}^{(0/1)}(s)$ by not introducing outer functions in $s$, but observing that if $\Omega_l(s) \rightarrow 1/s$ and $\tilde{d}_{l,i}^{(0/1)}(s)$ is at most constant for $s \rightarrow \infty$, the $s$-integral is finite and we can expand
\begin{align}
    \bar{d}_{l,i}^{(I)}(s) = \sum_j c^{(I)}_{l,ij} q_{l,j}(y_s)\,,
\end{align}
where the $q_{l,j}$ are constructed such that the imaginary part at the inelastic threshold grows like $(s - s_\text{in})^{l+1/2}$, as required for two-particle inelasticities. For $l = 0$, these are simply the monomials $y_s$, whereas for $l = 1$, the appropriate polynomials are constructed in Ref.~\cite{Bourrely:2008za}. For $l = 2$, the polynomials are given by
\begin{align}
    q_{2,j}(y) = y^j + \frac{(-1)^{j-N}j}{3+2N}\Bigg(\frac{(N+2)^2 - j^2}{N+1} y^{N+1}\nonumber\\ + \frac{(N+1)^2 - j^2}{N+2}y^{N+2}\Bigg)\,,
\end{align}
where $N$ is the truncation order.
The corresponding unitarity bound is not diagonal in the expansion coefficients, but can be computed through numerical integration. A slight modification of the form factors $f$ and $g$ is required, as can be seen from the additional factors of $s$ and $t$ in the numerator of Eq.~\eqref{eq::unitarity2}. These require that the corresponding form factors are expanded as
\begin{align}
    \bar{d}_{l,i}^{(I)}(s) = \frac{1}{\sqrt{s}}\sum_j c^{(I)}_{l,ij} q_{l,j}(y_s)\,.
\end{align}
Aside from the correct threshold behavior, this parameterization has the advantage that the Omnès factor is not canceled by the corresponding outer function above the inelastic threshold. Consequently, less terms in the expansion are required if the input phase describes the decay well.

\section{The di-pion invariant-mass spectrum}\label{sec::SPD}
The factorization of the $B\rightarrow\pi\pi\ell\nu$ form factors into Omnès function and a remainder encoding the $q^2$-dependence and inelastic effects allows us to exploit the available high-precision information on the $\pi\pi$ scattering phase shifts. These have been determined precisely using the constraints from Roy (and similar) equations~\cite{Roy:1971tc} and low-energy $\pi\pi$ scattering measurements as well as, crucially for the S-wave, $K\rightarrow\pi\pi\ell\nu$ decays~\cite{Ananthanarayan:2000ht,Garcia-Martin:2011iqs,Caprini:2011ky}.

For our study in Sec.~\ref{sec::Belle}, we restrain our partial-wave expansion to the S-, P- and D-waves. F-waves and higher are highly phase-space suppressed and show limited phase-motion at low $s$. Furthermore, the first F-wave resonance, the $\rho_3(1690)$ is highly inelastic and located above our region of interest.

In the following, we summarize the relevant knowledge on the three partial waves under consideration and their treatment in our analysis.

\subsection{S-wave}
The isoscalar S-wave is the major source of non-P-wave $B\rightarrow\pi\pi\ell\nu$ decays in the $\rho$ region. Consequently, to determine the P-wave fraction precisely, a reliable description of the S-wave contribution is crucial.
Two poles appear in the S-wave at energies below $1\GeV$: the $f_0(500)$ at $\sqrt{s} = (400 - 550) - i (200 - 350)\MeV$ and the $f_0(980)$ at $\sqrt{s} = (980 - 1010) - i (20 - 35)\MeV$, near the $K\bar{K}$ threshold~\cite{ParticleDataGroup:2024cfk}. Although the pole of the $f_0(500)$ sits far in the complex plane and is often quoted with a large uncertainty, advanced dispersive analyses that do not use Breit--Wigner-like lineshapes narrow the position down to $\sqrt{s} = 449^{+22}_{-16} - i (275 \pm 12)\MeV$~\cite{Pelaez:2015qba}. For a more detailed discussion on the pole determinations, see the review on \textit{Scalar Mesons below $1 \GeV$} in Ref.~\cite{ParticleDataGroup:2024cfk}.
The lineshape resulting from the interplay of the two poles gets further complicated by the onset of large inelasticities due to the $K\bar{K}$ channel, resulting in a dip, rather than a peak, near the $f_0(980)$ in processes with a light-quark source. In contrast, in neutral current decays of $B_s$ mesons a narrow peak is observed, rather than a dip~\cite{LHCb:2014ooi,LHCb:2014yov}.

As a consequence, the S-wave cannot strictly be treated as a single-channel problem and we must include the inelasticities due to the $K\bar{K}$ channel. To this end, we employ the results of Refs.~\cite{Daub:2015xja,Ropertz:2018stk}, where the solutions to the Roy equations of Refs.~\cite{Ananthanarayan:2000ht,Caprini:2011ky} are combined with S-wave $\pi\pi\rightarrow K\bar{K}$ scattering data~\cite{Cohen:1980cq,Etkin:1981sg,Buettiker:2003pp} and angular analysis of $B_{(s)}\rightarrow J/\Psi\pi\pi$ decays by LHCb~\cite{LHCb:2014ooi,LHCb:2014vbo}. The resulting two-channel Omnès matrix can be converted into an effective single-channel function, which has the correct elastic $\pi\pi$ S-wave scattering phase below the $K\bar{K}$ threshold, but follows the lineshape of $B\rightarrow J/\Psi\pi\pi$ decays at higher di-pion invariant masses. The largest uncertainty of the lineshape is the pion-to-kaon ratio when converting from the two-channel to the single-channel case and is controlled by one parameter: $r_K = \Gamma^n_K(0)/\Gamma^n_\pi(0)$, where $\Gamma^n_K(0)$ and $\Gamma^n_\pi(0)$ are the light-flavor pion and kaon scalar form factors at $s = 0$~\cite{Daub:2015xja}. While this treatment is similar to Ref.~\cite{Kang:2013jaa} it introduces a crucial improvement. In general the $B\rightarrow J/\Psi\pi\pi$ and $B\rightarrow\pi\pi\ell\nu$ S-wave form factors do not share the same phase above the $K\bar{K}$ threshold. However, the $y$-expansion develops additional phases above $s_\text{in}^{(0)} = 4 M_K^2$ and, consequently, will account for any difference.

\subsection{P-wave}
Given the prominence of the $\rho^0$-peak observed in the Belle analysis~\cite{Belle:2020xgu}, having good control over the isovector P-wave lineshape is paramount. To this end, we employ the high-precision determination of the P-wave phase shift obtained in Ref.~\cite{Colangelo:2018mtw}. In this work, the results of the Roy analysis of Refs.~\cite{Ananthanarayan:2000ht,Caprini:2011ky} are further constrained by data on the space- and time-like pion vector form factor from F2~\cite{Dally:1982zk} and NA7~\cite{NA7:1986vav}, as well as SND~\cite{Achasov:2005rg,Achasov:2006vp}, CMD-2~\cite{CMD-2:2001ski,CMD-2:2003gqi,Aulchenko:2006dxz,CMD-2:2006gxt}, BaBar~\cite{BaBar:2009wpw,BaBar:2012bdw}, and KLOE~\cite{KLOE:2008fmq,KLOE:2010qei,KLOE:2012anl,KLOE-2:2017fda}, respectively. The resulting phase shift has negligible uncertainty in the (quasi-)elastic region below the $\pi\omega$ threshold and, consequently, we neglect it in our analysis. Uncertainties in the region above the $\pi\omega$ threshold can also be ignored, as any deviation from the elastic P-wave phase shift is absorbed in the $y$-expansion of the P-wave form factors, similar to the S-wave case.

While mixing between the isovector and isoscalar P-wave is only induced through small isospin-breaking effects, it is enhanced in the region around the $\rho$-peak due to the small mass difference between the $\rho$ and the $\omega$. As the $\omega$ is very narrow, the effect of $\rho$--$\omega$ mixing can be included by replacing the isovector P-wave Omnès factor according to $\Omega_1^1(s) \rightarrow G_\omega(s)\Omega_1^1(s)$~\cite{Gardner:1997ie,Leutwyler:2002hm}, where
\begin{align}
    G_\omega(s) &= 1 + \frac{s}{\pi}\int_{9M_\pi^2}^\infty\mathrm{d}s^\prime\frac{\mathrm{Im}g_\omega(s^\prime)}{s^\prime(s^\prime-s)}\left(\frac{1-\frac{9M_\pi^2}{s^\prime}}{1-\frac{9M_\pi^2}{M_\omega^2}}\right)^4\,,\notag\\
    g_\omega(s) &= 1 + \epsilon_\omega \frac{s}{(M_\omega - \frac{i}{2}\Gamma_\omega)^2 - s}\,.
\end{align}
Here, $G_\omega$ has the correct analytic structure at the $3\pi$ threshold, while it is real below and $\epsilon_\omega$ is a real parameter. In general, $\epsilon_\omega$ acquires a small imaginary part through the presence of the $\pi^0\gamma$ and other radiative channels~\cite{Colangelo:2022prz}. While straightforward to include, at the current level of precision of the available data on $B\rightarrow\pi\pi\ell\nu$ decays, this effect cannot be resolved. The constant $\epsilon_\omega$ determined from the pion vector form factor in Ref.~\cite{Colangelo:2018mtw} cannot directly be applied to $B\rightarrow\pi\pi\ell\nu$ decays.
However, following Refs.~\cite{Daub:2015xja,Ropertz:2018stk,Holz:2022hwz}, we rescale $\epsilon_\omega$ by the relative isoscalar-to-isovector ratio between $B\rightarrow\pi\pi\ell\nu$ decays and the pion vector form factor. To this end, we decompose the relevant quark currents:
\begin{align}
    j^\mu_\text{EM} &= \frac{1}{2}(\bar{u}\gamma^\mu u - \bar{d}\gamma^\mu d) + \frac{1}{6}(\bar{u}\gamma^\mu u + \bar{d}\gamma^\mu d)\,,\notag\\
    \bar{u}\gamma^\mu u &= \frac{1}{2}(\bar{u}\gamma^\mu u - \bar{d}\gamma^\mu d) + \frac{1}{2}(\bar{u}\gamma^\mu u + \bar{d}\gamma^\mu d)\,.
\end{align}
Evidently, the isoscalar-to-isovector ratio is a factor of 3 greater than for the electromagnetic current and we obtain $\epsilon_\omega = 3 \epsilon_\omega^{\text{EM}}$. As a consequence, the sharp edge seen in the pion vector form factor will also occur in $B\rightarrow\pi\pi\ell\nu$ decays, but will be further enhanced.

At higher invariant masses, the $\rho^\prime$ and $\rho^{\prime\prime}$ resonances contribute, yet they predominantly decay to four pions. In principle they could be approximated by continuing the P-wave phase shift appropriately in the inelastic regime~\cite{Schneider:2012ez}, but it is unclear how reliable this procedure is in the case at hand.
Instead, the $y$-expansion is able to account for effects due to the higher resonances, should the data require it.

\subsection{D-wave}
Compared to the other two partial waves, the isoscalar $\pi\pi$ D-wave is relatively simple.
In the $\rho$ region the D-wave phase increases slowly, but above the $K\bar{K}$ threshold the phase motion becomes significant and crosses through $\pi/2$ around $\sqrt{s} \approx 1270\MeV$. This fast motion is associated with the lightest tensor resonance, the $f_2(1270)$ with a well-determined pole location at $\sqrt{s} = (1260 - 1283) - i (90 - 110)\MeV$~\cite{ParticleDataGroup:2024cfk}. The $f_2(1270)$ lineshape is generally well described by a Breit--Wigner and is largely elastic, i.e., it dominantly couples to the $\pi\pi$ final state. This can be understood from the D-wave suppression of inelasticities, which can only grow as $(s-4 M_K^2)^{5/2}$ near threshold. Furthermore, the next isoscalar tensor resonance, the $f^\prime_2(1525)$, dominantly couples to kaons and $s\bar{s}$ sources and we do not expect it to contribute to $B\rightarrow\pi\pi\ell\nu$ decays in any significant manner.

While the $f_2(1270)$ is located in the inelastic region, we take the elastic D-wave phase shift from Ref.~\cite{Garcia-Martin:2011iqs} and absorb any deviations in the $y$-expansion.\footnote{Recently, Ref.~\cite{Pelaez:2024uav} improving the description of the D-wave appeared, allowing for further improvements of our work in the future.} However, given the largely elastic nature of the $f_2(1270)$, the suppressed onset of inelasticities and the absence of nearby resonances that couple to the $\pi\pi$ final state, this treatment results in an accurate description of the D-wave lineshapes for energies up to $\sqrt{s} \approx 1.5\GeV$.

In the Belle measurements of Refs.~\cite{Belle:2013hlo,Belle:2020xgu} a resonant structure was observed in the region where the $f_2(1270)$ is expected. However, neither analysis could unambiguously establish the existence of $B^+\rightarrow f_2(1270)\ell^+\nu$ decays due to the lack of control over the S- and P-wave in this energy region. With our model-independent parameterization we are able to study the resonant structure seen in Ref.~\cite{Belle:2020xgu} and determine if it is caused by the $f_2(1270)$.

\section{Fit to Belle data}\label{sec::Belle}
The analysis in Ref.~\cite{Belle:2020xgu} was carried out using a hadronic tagged reconstruction approach with the complete Belle data set comprising a total integrated luminosity of $\unit[711]{fb^{-1}}$, collected by the Belle detector at the $\Upsilon(4S)$ resonance. While the main result is the measurement of the total branching fraction of $B^{+} \rightarrow \pi^{+}\pi^{-} \ell^+ \nu$ decays, the partial branching fractions are also provided in bins of $M_{\pi\pi}$, $q^{2}$, as well as a two-dimensional analysis in $13$ bins of both variables. These results are unfolded to correct for detector resolution and acceptance effects. The analysis did not consider a non-resonant inclusive component, since the simulated contributions from high-multiplicity mass modes such as $B^{+} \rightarrow \pi^{+}\pi^{-}\pi^{0} \ell^+ \nu$ and $B^{+} \rightarrow \pi^{+}\pi^{-}\pi^{0} \pi^{0} \ell^+ \nu$ decays were found to be negligible after the full selection criteria were applied. The most significant source of systematic uncertainty was due to the modeling of signal processes and the lack of precise knowledge of hadronic form factors describing specific exclusive decay modes, ranging between 4.46\% to 29.9\% in different bins of $M_{\pi\pi}$ and $q^{2}$ for the two-dimensional fit scenario.

Given the large uncertainty and coarse binning of the data, especially at low $M_{\pi\pi}$, crossed-channel effects, such as enhancements from the $B^\ast$ pole, cannot be resolved at present. Thus, we focus on the $s$-channel only and neglect the isospin $1/2$ and $3/2$ $t$- and $u$-channel contributions. Once more precise data becomes available, the full formalism of Sec.~\ref{sec::reco} can be applied.

We perform a Bayesian fit to the $2$D spectrum with the EOS software~\cite{EOSAuthors:2021xpv}, in which we implemented our parameterization. We employ the version of the parameterization implementing the correct scaling at the inelastic threshold at the price of a more complex form of the unitarity bound, choosing $q^2_0 = 0\GeV^2$ and $s_0 = 4 M_\pi^2$. The inelastic thresholds are set to $\sqrt{s_\text{in}} = (M_\omega + M_\pi)$ for the P-wave, as well as $\sqrt{s_\text{in}} = 2 M_K$ for the S- and D-waves.

In a first step we scan the range of the expansion coefficients allowed by unitarity to increase the efficiency during the actual fit. For the fit itself we use uniform priors for the expansion coefficients, as well as an uniform prior for $r_K$ between $0.4$ and $0.6$~\cite{Daub:2015xja}. Since we do not have external constraints on the form factors, our fit is insensitive to $|V_{ub}|$ and we take the default values in EOS, $|V_{ub}| = 0.0036$. To implement the unitarity bounds we follow Ref.~\cite{Bordone:2019vic} and implement a penalty term in the likelihood. We take $\chi_T^{(A)} = \chi_T^{(V)} = 5.742\times 10^{-4}\GeV^{-2}$, obtained from the three-loop calculation of Ref.~\cite{Hoff:2011ge}, and assign an uncertainty of $5\%$. For the vector current we subtract the contribution of the $B^\ast$ subthreshold resonance, which amounts to $9\%$ of the bound.

To study different truncations of the $y$-$z$-expansion, we choose three different fit scenarios.
In the simplest scenario, the $2/0/0$ scenario,  we terminate the $y$-expansion at leading order, but include two terms in the $z$-expansion for each form factor. In the $2/1/0$ scenario we have two terms in the $z$-expansion at $\mathcal{O}(y^0)$ and one at $\mathcal{O}(y^1)$, while for the $3/2/1$ scenario we have three at $\mathcal{O}(y^0)$, two at $\mathcal{O}(y^1)$ and one at $\mathcal{O}(y^2)$. The three scenarios have $15$, $22$, and $43$ parameters, compared to the $13$ data points, and result in a perfect fits. In the following, we present our results for the $3/2/1$ fit scenario. Results for the other two scenarios can be found in Appendix~\ref{sec::further_fits}.

\subsection{Partial branching ratios and \texorpdfstring{$M_{\pi\pi}$}{Mππ} spectrum}
Because of the distinct lineshapes of the individual partial waves, the fit to Belle data effectively distinguishes the different contributions. For the branching fractions, we obtain
\begin{align}
    \mathcal{B}(B^+\rightarrow (\pi^+ \pi^-)_S \ell^+ \nu) &= 2.2^{+1.4}_{-1.0}\times 10^{-5}\,,\nonumber\\
    \mathcal{B}(B^+\rightarrow (\pi^+ \pi^-)_P \ell^+ \nu) &= 19.6^{+2.8}_{-2.7}\times 10^{-5}\,,\nonumber\\
    \mathcal{B}(B^+\rightarrow (\pi^+ \pi^-)_D \ell^+ \nu) &= 3.5^{+1.3}_{-1.1}\times 10^{-5}\,.\label{eq::fit1}
\end{align}
The correlations between the different contributions are small and the D-wave is more than $2\sigma$ away from $0$.

\begin{figure}
    \centering
    \includegraphics[width=\linewidth]{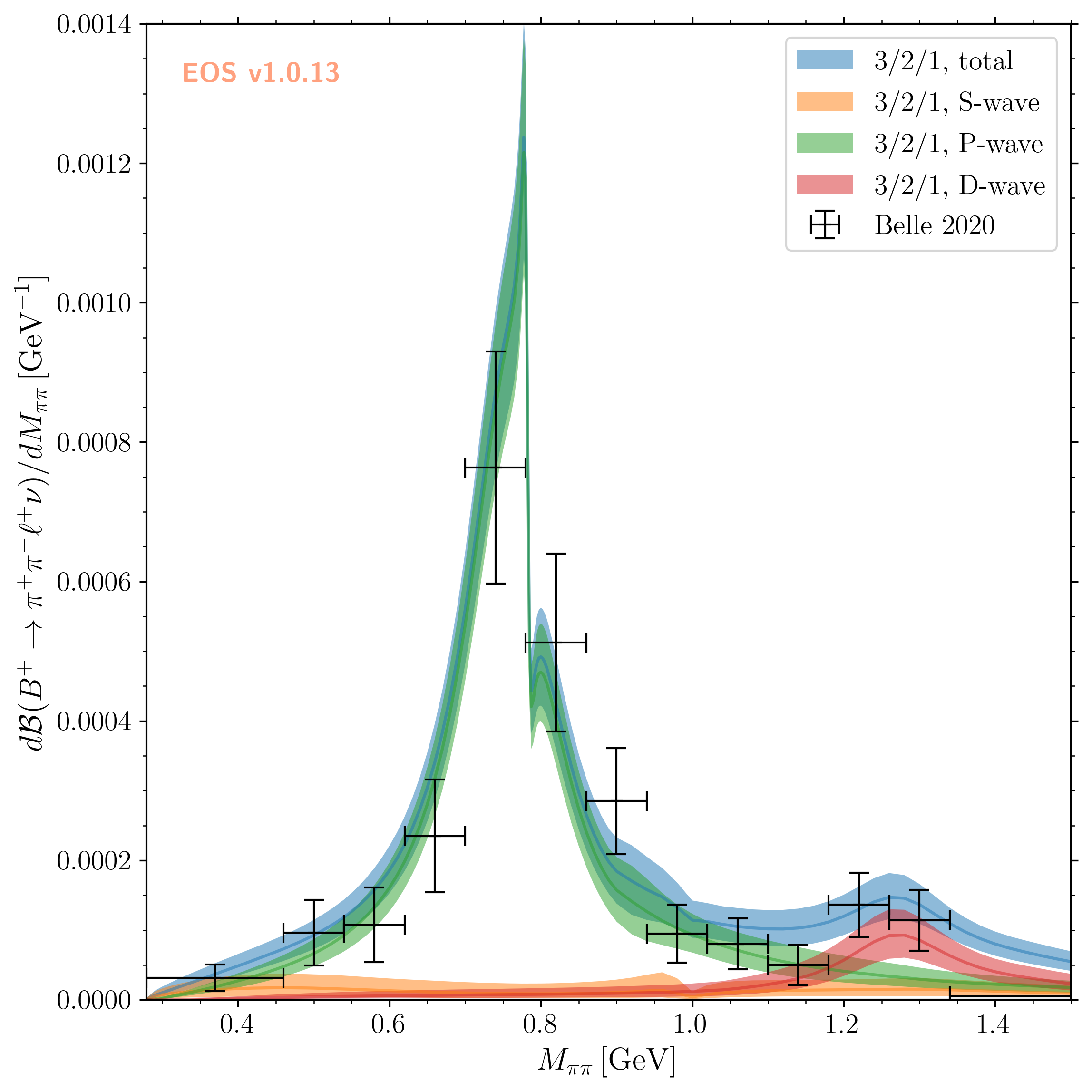}
    \caption{The $B^+\rightarrow \pi^+ \pi^- \ell^+ \nu$ $M_{\pi\pi}$ spectrum. The different bands show the contributions of the three partial waves as well as their sum. The data points are from the $1$D-measurement of Ref.~\cite{Belle:2020xgu}.}
    \label{fig:mpipi_charged}
\end{figure}

In Fig.~\ref{fig:mpipi_charged} we present our results for the $B^+\rightarrow \pi^+ \pi^- \ell^+ \nu$ $M_{\pi\pi}$ spectrum and compare it to the $1$D-measurement of Ref.~\cite{Belle:2020xgu}. Note that coarser $2$D-data is used in the fit, yet we find excellent agreement with the finer binned $1$D-data.
At low invariant masses, near the threshold, the S- and P-wave are of similar size, while the D-wave is completely negligible, as expected. In the region around the $\rho$-peak there is only a small contamination from the S- and D-waves, while the $\rho$--$\omega$ mixing leads to a significant distortion of the lineshape. At current precision, this structure is not resolved by the available data, but measurements at Belle II and LHCb should be able to observe such a drastic feature. Near the P-wave inelastic threshold at $(M_\omega + M_\pi)$ there is a significant increase in the uncertainty on the P-wave contribution, demonstrating the power of our parameterization: despite making assumptions on the lineshape also above the inelastic threshold, the higher terms in the $y$-expansion smear them out if the data allows. Similarly, the S-wave uncertainty grows near the $K\bar{K}$ threshold, overshooting the uncertainty due to the limited knowledge of $r_K$.
Above $1\GeV$ the D-wave becomes relevant and exhibits a Breit--Wigner-like peak for the $f_2(1270)$. In the $\pi^+\pi^-$ mode there is a sizeable background from the P- and the S-wave in this region, complicating the extraction of the D-wave component without the use of additional angular information.

It is interesting to compare our results to those reported by Ref.~\cite{Bernlochner:2024ehc}, where a fit to the $1$D $M_{\pi\pi}$ spectrum of Ref.~\cite{Belle:2020xgu} below $1.02\GeV$ is performed using a resonance model for the S- and P-waves. Reference~\cite{Bernlochner:2024ehc} quotes $\Delta\mathcal{B}(B^+\rightarrow (\pi^+ \pi^-)_S \ell^+ \nu) < 5.1 \times 10^{-5}$ at 90\% confidence level for invariant masses below $1.02\GeV$, as well as $\mathcal{B}(B^+\rightarrow\rho^0\ell\nu) = 14.1^{+4.9}_{-3.8} \times 10^{-5}$. We can directly compare the upper limit on the S-wave, for which we obtain $\Delta\mathcal{B}(B^+\rightarrow (\pi^+ \pi^-)_S \ell^+ \nu) < 2.4 \times 10^{-5}$ at 90\% confidence level, an improvement by more than a factor of $2$. A direct comparison for the P-wave is not possible. While Ref.~\cite{Bernlochner:2024ehc} also includes $\rho$--$\omega$ mixing, albeit without resorting to the methods used here, it is unclear up to which value in $M_{\pi\pi}$ it can directly be compared to our P-wave results. However, when comparing the uncertainty of Ref.~\cite{Bernlochner:2024ehc} to our results in Eq.~\eqref{eq::fit1} we find a significant improvement.

\begin{figure}
    \centering
    \includegraphics[width=\linewidth]{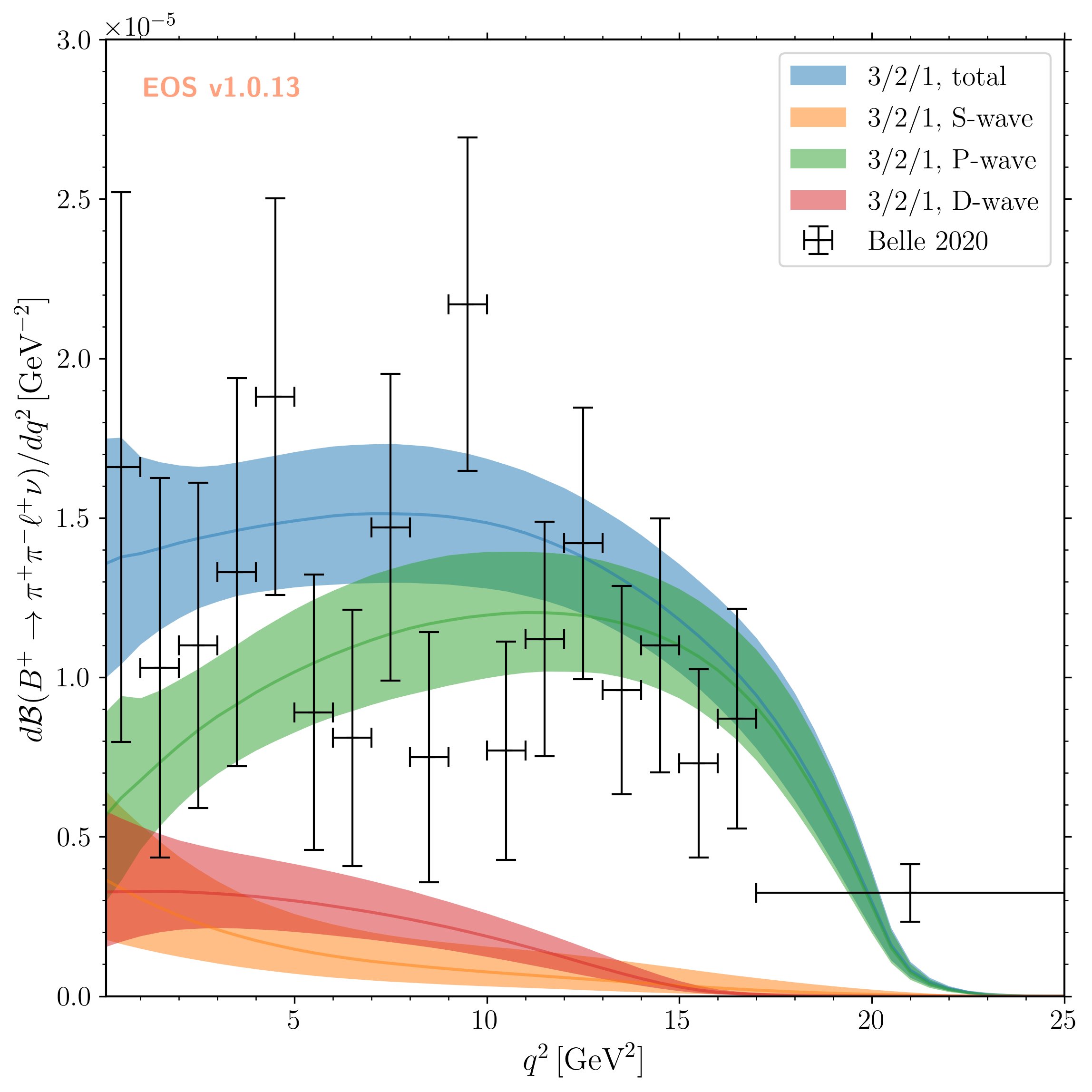}
    \caption{The $B^+\rightarrow \pi^+ \pi^- \ell^+ \nu$ $q^2$ spectrum. The different bands show the contributions of the three partial waves as well as their sum. The data points are from the $1$D-measurement of Ref.~\cite{Belle:2020xgu}.}
    \label{fig:q2_charged}
\end{figure}

\subsection{\texorpdfstring{$q^2$}{q²}-dependence and saturation of the unitarity bounds}
In Fig.~\ref{fig:q2_charged} we compare the $q^2$ spectrum we obtain from the fit to the $2$D-measurement of Ref.~\cite{Belle:2020xgu} to the respective $1$D-measurement. Overall, the agreement is excellent and the uncertainties on the obtained spectra are under control.

While the high-$q^2$ region is saturated by the P-wave contribution, the other two partial waves contribute significantly below $12\GeV^2$. At $q^2 = 0\,\GeV^2$, the sum of S- and D-waves is approximately of the same size as the P-wave contribution. The P-wave and, to a lesser extent the D-wave, quickly rise below the kinematic endpoint, which is due to the presence of the $B^\ast$ pole in the vector form factor. This behavior is similar to the case of $B\rightarrow D\pi\ell\nu$ decays, where the $B_c^\ast$ pole in the vector form factor is the closest to the kinematic region~\cite{Gustafson:2023lrz}. Furthermore, in contrast to a narrow-width treatment of the $\rho$, the P-wave spectrum extends to $q^2$-values beyond $(M_B - M_\rho)^2 \approx 20.3\GeV^2$.

\begin{figure}
    \centering
    \includegraphics[width=\linewidth]{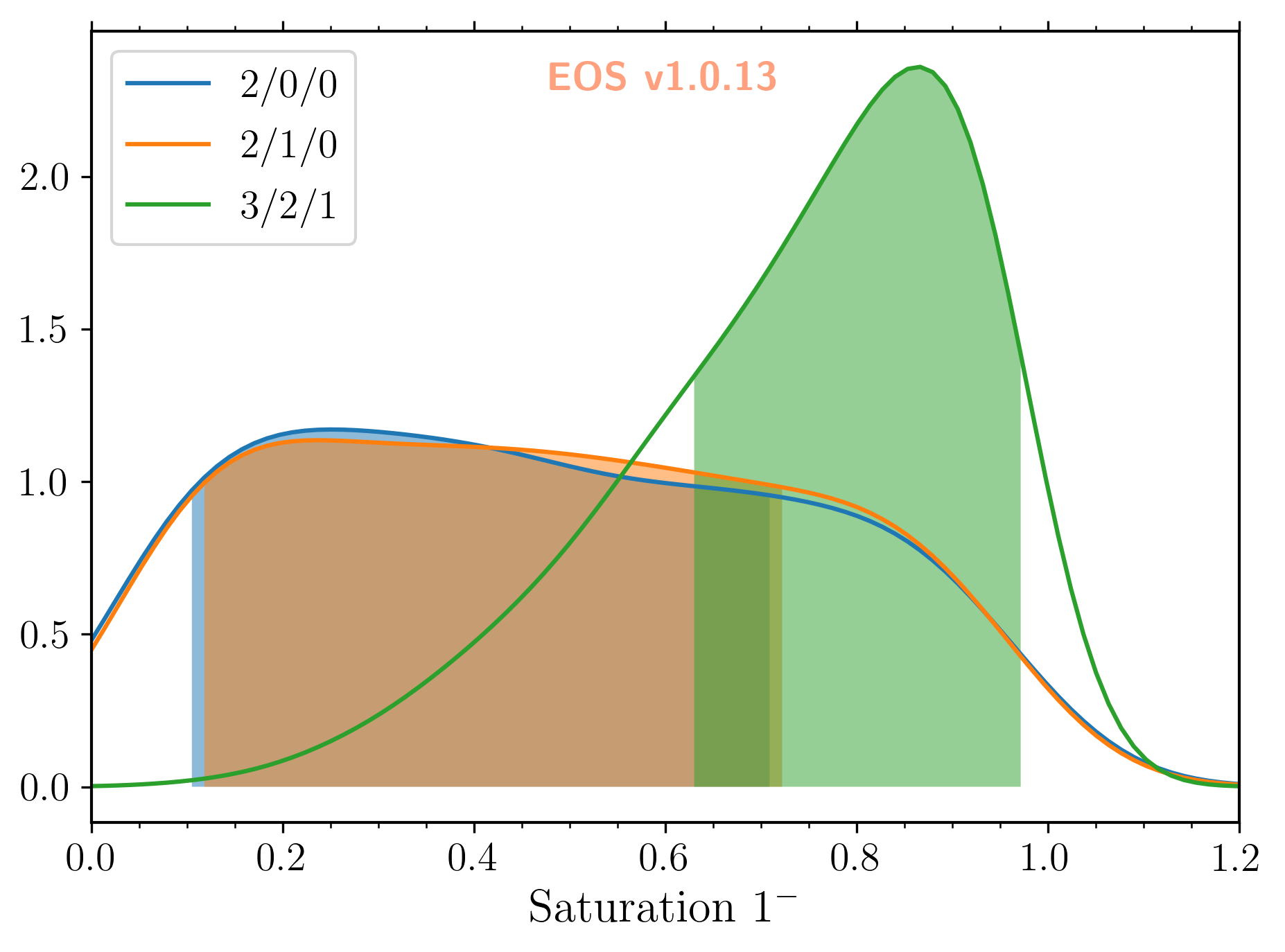}
    \caption{Saturation of the $1^-$ unitarity bound due to the $B\rightarrow \pi\pi$ form factors for the three different fit scenarios discussed in the text. The shaded regions correspond to $68\%$ confidence intervals. The scale of the $y$-axis is given in arbitrary units.}
    \label{fig:sat_1m}
\end{figure}

\begin{figure}
    \centering
    \includegraphics[width=\linewidth]{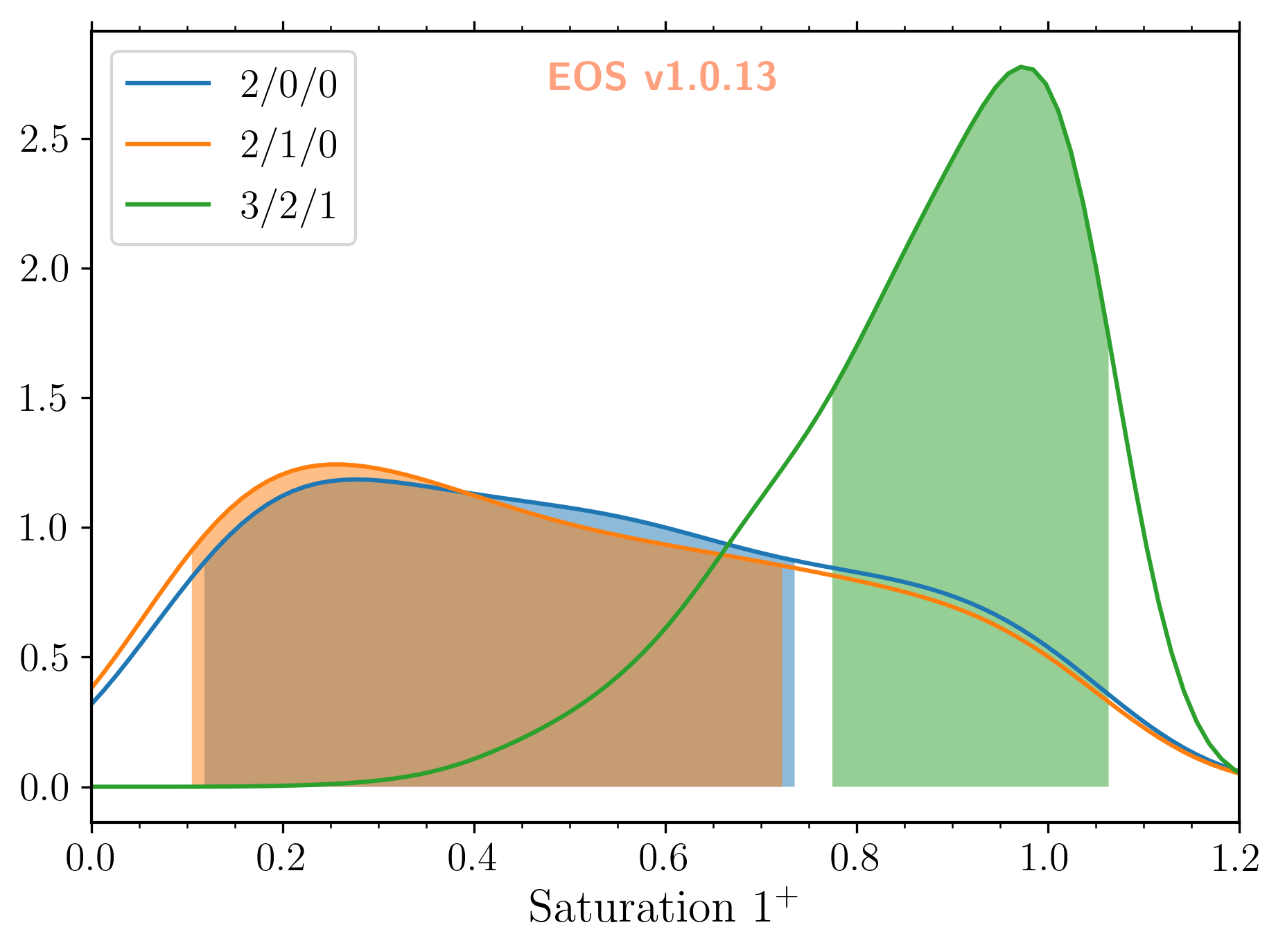}
    \caption{Saturation of the $1^+$ unitarity bound due to the $B\rightarrow \pi\pi$ form factors for the three different fit scenarios discussed in the text. The shaded regions correspond to $68\%$ confidence intervals. The scale of the $y$-axis is given in arbitrary units.}
    \label{fig:sat_1p}
\end{figure}

Figures~\ref{fig:sat_1m} and \ref{fig:sat_1p} show the saturation of the unitarity bounds in the three different fit scenarios, which is calculated by determining the right-hand side of the relevant versions of Eq.~\eqref{eq::bound1} for the transverse components of the vector and axial currents. While the saturation remains largely unaffected by increasing the expansion order from $2/0/0$ to $2/1/0$, the inclusion of more terms in the $z$-expansion for the $3/2/1$ scenario increases the saturation significantly. A similar behavior is observed in the fits of Ref.~\cite{Gubernari:2023puw} for form factors where only LCSR calculations at low $q^2$, but no lattice-QCD (LQCD) results at high $q^2$ are available. In this case the saturation increases with increasing truncation order of the $z$-expansion and finally peaks near $1$. Consequently, the unitarity bounds are saturated and given that $|z| < 1$ in the decay region, increasing the truncation order does not change the resulting form factors.
While we are not dealing with an extrapolation here, as the Belle data covers both low- and high-$q^2$ regions, the coarse binning in $q^2$ and large uncertainties leave significant freedom for the expansion coefficients and thus we observe a similar behavior. The main effect of the unitarity bounds in this work is the restriction of the allowed shape of the $q^2$ spectrum, as can be seen from the significantly smaller uncertainty of the $q^2$ spectrum that we obtain, compared to the measured spectrum provided by Belle.

The slightly lower saturation of the $1^-$ saturation compared to the $1^+$ saturation is due to the $B^\ast$ contribution to the $1^-$ unitarity bound. The uncertainties on the susceptibilities $\chi_T^{(A)}$ and $\chi_T^{(V)}$ smear out of the otherwise sharp edge at $1$.

Given the relatively large saturation, a global fit of $b\rightarrow u$ form factors extending the one performed in Ref.~\cite{Leljak:2023gna} could result in reduced uncertainties for less well known form factors, such as those for $B\rightarrow\eta\ell\nu$, $B\rightarrow\eta^\prime\ell\nu$, or $B\rightarrow\omega\ell\nu$ decays.

\subsection{Predictions for \texorpdfstring{$B^+\rightarrow \pi^0\pi^0\ell^+\nu$}{B⁺→π⁰π⁰l⁺ν} decays}
\begin{figure}[h]
    \centering
    \includegraphics[width=\linewidth]{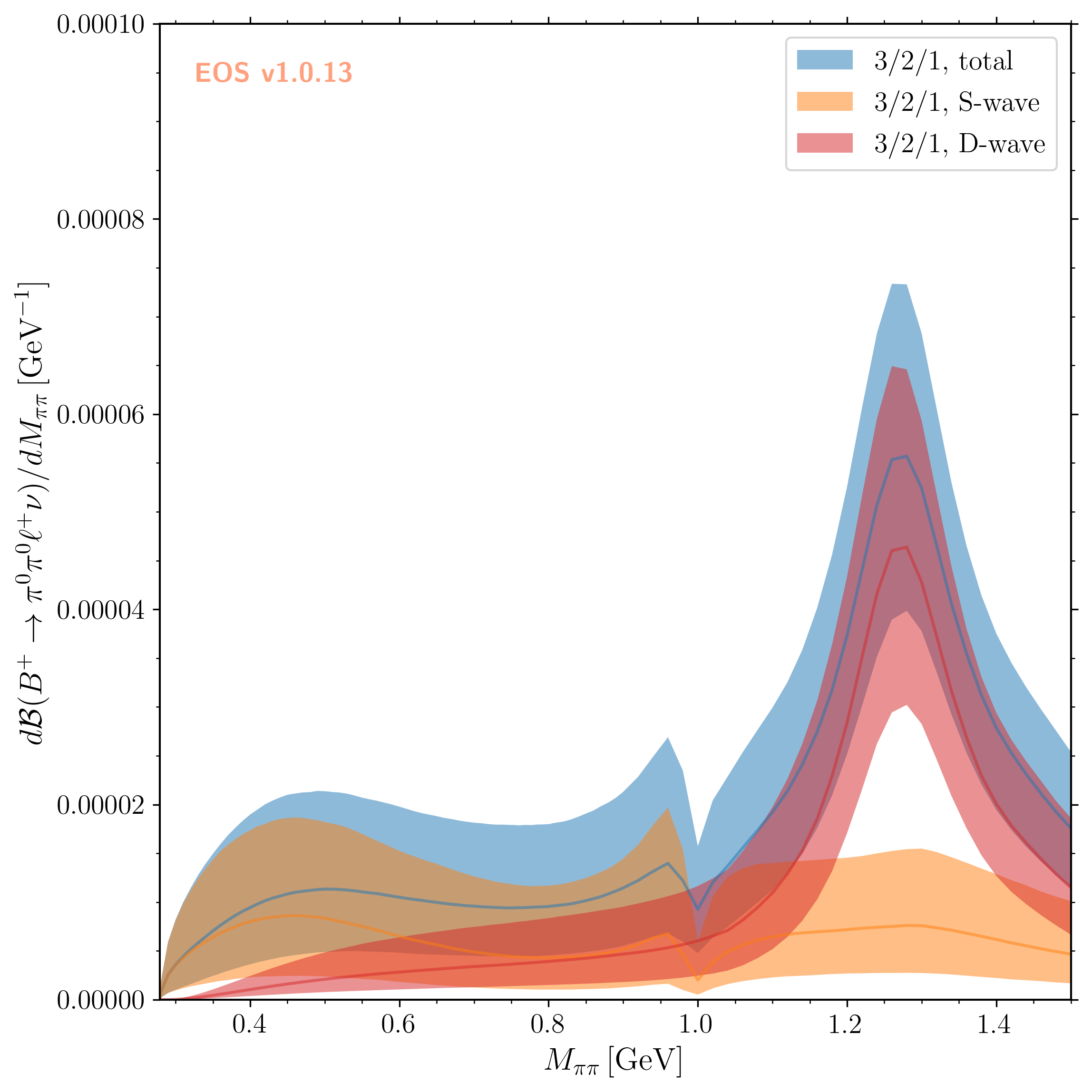}
    \caption{The $B^+\rightarrow \pi^0 \pi^0 \ell^+ \nu$ $M_{\pi\pi}$ spectrum. The different bands show the contributions of the two contributing partial waves as well as their sum.}
    \label{fig:mpipi_neutral}
\end{figure}

\begin{figure}[H]
    \centering
    \includegraphics[width=\linewidth]{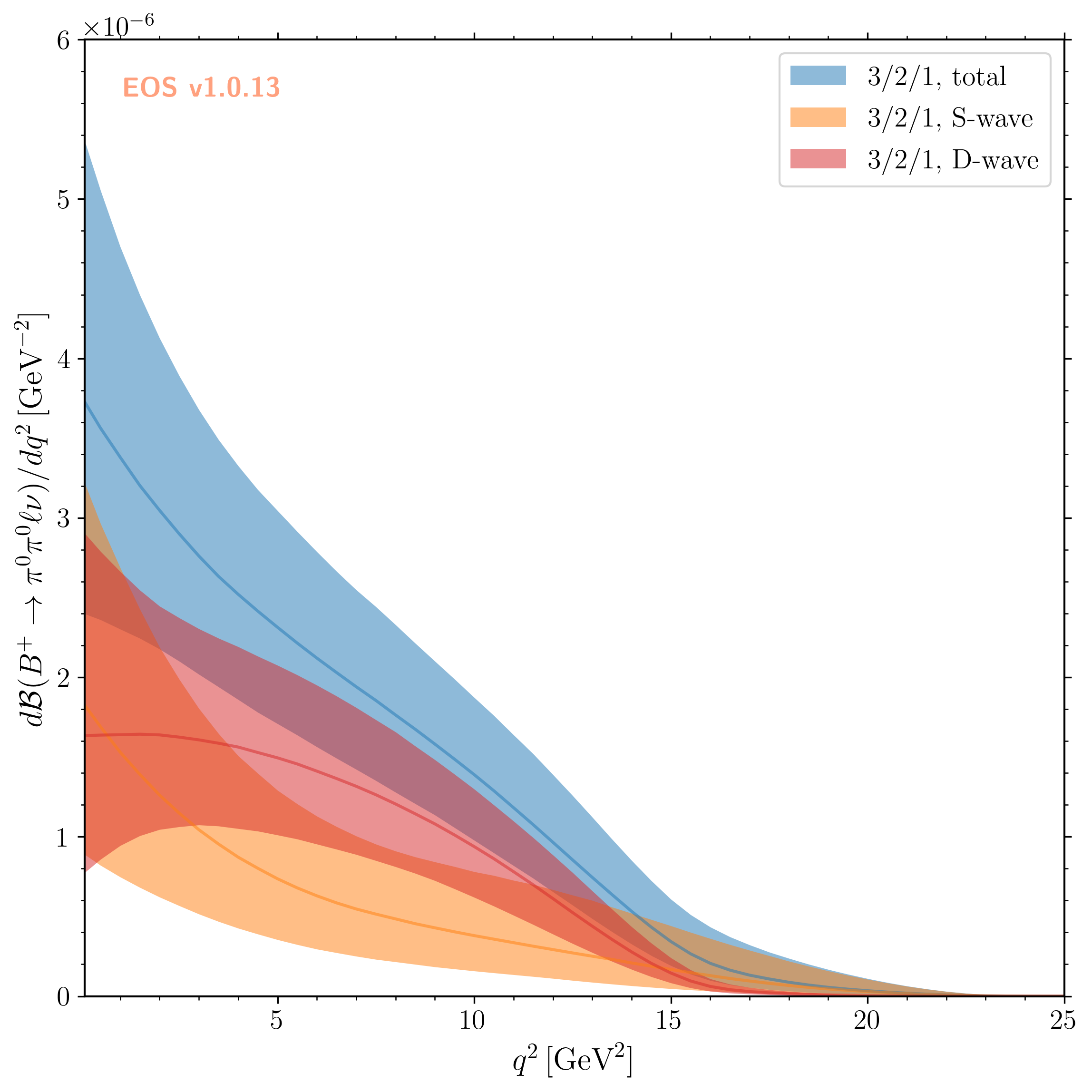}
    \caption{The $B^+\rightarrow \pi^0 \pi^0 \ell^+ \nu$ $q^2$ spectrum. The different bands show the contributions of the two contributing partial waves as well as their sum.}
    \label{fig:q2_neutral}
\end{figure}

With our results for $B^+\rightarrow \pi^+\pi^-\ell^+\nu$ decays at hand, we can obtain predictions for the yet unobserved $B^+\rightarrow \pi^0\pi^0\ell^+\nu$ decay. Only the S- and D-waves contribute, as the $\pi^0\pi^0$ system is always in an isoscalar configuration, resulting in a sizeable branching fraction of
\begin{align}
    \mathcal{B}(B^+\rightarrow \pi^0\pi^0\ell^+\nu) = 2.9^{+0.9}_{-0.7}\times 10^{-5}\,,
\end{align} 
comparable to the $B^+\rightarrow \eta \ell^+ \nu$ and $B^+\rightarrow \eta^\prime \ell^+ \nu$ branching fractions~\cite{ParticleDataGroup:2024cfk}.
While the relative precision is still limited, it is similar in size to that of the $B\rightarrow \eta^\prime \ell \nu$ branching fraction.
In addition, we obtain the $M_{\pi\pi}$- and $q^2$-dependence of the decay rate, shown in Figs.~\ref{fig:mpipi_neutral} and \ref{fig:q2_neutral}, respectively.

The absence of the P-wave contribution leads to a clearly visible $f_2(1270)$ peak in the $M_{\pi\pi}$ spectrum, in contrast to the $\pi^+\pi^-$ mode, making the $\pi^0\pi^0$ channel a promising, yet experimentally difficult, discovery channel for $B^+\rightarrow f_2(1270)\ell^+\nu$ decays. The region of $M_{\pi\pi} < 1\GeV$ is dominated by the S-wave and measurements in this region will help to establish the size without resorting to angular information.
The $q^2$ spectrum falls off towards high $q^2$, but a sizeable contribution remains in the region beyond the $B\rightarrow X_c \ell\nu$ endpoint.

\section{Implications and Outlook}\label{sec::Outlook}
The model-independent form factor parameterization introduced here allows us, for the first time, to extract the contributions of different partial waves to $B\rightarrow \pi\pi\ell\nu$ decays. The branching fractions obtained in Eq.~\eqref{eq::fit1} together with the $M_{\pi\pi}$ spectrum shown in Fig.~\ref{fig:mpipi_charged} allow for an assessment of the discrepancy between the determinations of the $B\rightarrow \rho^0\ell\nu$ branching fractions obtained by BaBar, Belle, and Belle II: we find only moderate S- and D-wave components below the $\rho$-peak and, consequently, it is likely that the BaBar and Belle II measurements overestimate the \textit{non-resonant} $B\rightarrow\pi\pi\ell\nu$ background, lowering the observed $B^+\rightarrow \rho^0\ell^+\nu$ branching fraction.
Our P-wave branching fraction is somewhat larger, but compatible with the $B^+\rightarrow \rho^0\ell^+\nu$ branching fraction reported by Belle~\cite{Belle:2013hlo}.
We confirm the evidence for a second resonance in the $\pi^+\pi^-$ spectrum near $1.3\GeV$, corresponding to the D-wave $f_2(1270)$ resonance, at the $2\sigma$ level.

Isospin relations allow us to obtain predictions for $B^0\rightarrow \pi^- \pi^0 \ell^+ \nu$ and $B^+\rightarrow \pi^0\pi^0\ell^+\nu$ decays. Only odd partial waves contribute to the former and, consequently, it is almost entirely made up by P-wave contributions, i.e., $B^0\rightarrow \rho^-\ell^+\nu$ decays, at low invariant masses. Thus, there are negligible additional $\pi\pi$ contributions in the $\rho^-$ region. The latter is an experimentally challenging process, but the sizeable branching fraction we obtain shows that it can be of importance as a background to interesting measurements such as $B^+\rightarrow \gamma\ell^+\nu$ or a substantial signal component for inclusive $B\rightarrow X_u\ell\nu$ decays.
Our form factor parameterization and fit results will allow us to incorporate this component into future analyses, reducing uncertainties related to this mode.

To obtain competitive and theoretically clean determinations of $|V_{ub}|$ in $B\rightarrow \pi\pi\ell\nu$ decays, significant work is required, both from theory and experiment. On the theoretical side, determinations of the P-wave form factors beyond the narrow-width limit need to mature. To this end, the LCSR calculations of Refs.~\cite{Hambrock:2015aor,Cheng:2017sfk,Cheng:2017smj,Cheng:2025hxe} need to be revisited using the form factor parameterization presented here and combined with LQCD calculations, which have recently been calculated at unphysical pion masses~\cite{Leskovec:2025gsw}, but will become available at the physical point in the next years.
Furthermore, constraining the S- and D-waves through LCSR calculations is feasible; see Ref.~\cite{Descotes-Genon:2023ukb} for the $K\pi$ S-wave in $B\rightarrow K\pi\ell\ell$ decays.
Experimentally, it would be advantageous to study kinematic distributions beyond the $q^2$- and $M_{\pi\pi}$ spectra. A measurement of the asymmetry of the $\cos\theta_\pi$ spectrum as a function of $M_{\pi\pi}$ is directly sensitive to the interference between S- and P-wave. Given the knowledge of their relative phase, this would allow for an improved separation of the two components. Furthermore, an explicit incorporation of the lineshapes and form factors presented in this work directly into experimental analyses is paramount to the upcoming LHCb $B^+\rightarrow\rho^0\ell^+\nu$ measurement~\cite{Kirsebom:2023lba} and future measurements at Belle II. Using hadronization algorithms or simulations following phase-space distributions to obtain these two-body contributions otherwise leads to systematic uncertainties that are both large and difficult to assess.

Our results for $B^+\rightarrow \pi^0\pi^0\ell^+\nu$ present additional physics opportunities. While the uncertainties for the S- and D-waves are still sizeable, a measurement of the $B^+\rightarrow \pi^0\pi^0\ell^+\nu$ with a precision better than $25\%$ would already reduce the uncertainties on the two components. A measurement of partial branching fractions in the two regions $M_{\pi\pi} \in [ 2 M_\pi, 1.0\GeV]$ and $M_{\pi\pi} \in [1.0\GeV, 1.5\GeV]$ could effectively constrain the S-wave contribution and strengthen the evidence for $B\rightarrow f_2(1270)\ell\nu$ decays.

The parameterization presented here can be directly applied to processes of interest beyond $B\rightarrow \pi\pi\ell\nu$ decays. One promising process is the study of semileptonic $D\rightarrow K\pi\ell\nu$ decays, measured to high precision at BES III~\cite{BESIII:2015hty,BESIII:2018jjm,BESIII:2024xjf}. This would allow for an improved determination of the $K\pi$ scattering phase shifts, especially for the S-wave, in a similar manner to the S-wave $\pi\pi$ phase shift in $K\rightarrow\pi\pi\ell\nu$ decays~\cite{Colangelo:2015kha}. Furthermore, in this case the crossed-channel contributions cannot simply be neglected, given the precision of the available data, providing an ideal scenario to study their impact. In this context, alternative treatments of the left-hand cut, e.g., with conformal transformations similar to the ones discussed in Ref.~\cite{Gopal:2024mgb}, could be investigated. The determined phase shifts can then be used to improve the description of $B_s \rightarrow K\pi\ell\nu$ decays, a background to $|V_{ub}|$ determinations in $B_s\rightarrow K\ell\nu$ decays at LHCb~\cite{LHCb:2020ist}, as well as rare $B\rightarrow K\pi\ell\ell$ decays.

Further applications are also in reach: By extending our parameterization to the multi-channel case it will be applicable to S-wave $B\rightarrow D\pi\ell\nu$ decays, improving over Ref.~\cite{Gustafson:2023lrz}, and can control the uncertainty of the S-wave contributions in $B_s \rightarrow D K \ell \nu$ decays, relevant for future measurements at LHCb~\cite{DeCian:2023ezb}. This extension would also allow us to get a better handle on the uncertainty of the $\pi\pi$ S-wave above the kaon threshold. If measurements of $B\rightarrow K\bar{K}\ell\nu$ decays at Belle II and LHCb become available, a simultaneous study could be conducted, allowing for a better isolation of the $B\rightarrow f_0(980)\ell\nu$ contribution. However, the $K^+ K^-$-channel alone would be insufficient due to an admixture of isovector and isoscalar $K\bar{K}$ contributions. To this end, either the $K^0_S K^0_S$ or $K^\pm K^0_S$ final states need to be measured as well, or pure isovector $B\rightarrow \eta\pi\ell\nu$ decays need to be studied (cf.\ the analogous discussion for $B^0\to J/\Psi \{\pi\eta,K\bar{K}\}$ in Ref.~\cite{Albaladejo:2016mad}). These measurements would not only improve our understanding of isoscalar $B^+\rightarrow f_0(980)\ell^+\nu$ and $B^+\rightarrow f_2(1270)\ell^+\nu$ decays, but also their isovector relatives, $B\rightarrow a_0(980)\ell\nu$ and $B\rightarrow a_2(1320)\ell\nu$ decays.

While this work only presents a necessary first step into the study of semileptonic decays with two or more final-state hadrons, the results we obtained have far-reaching consequences for determinations of $|V_{ub}|$ in $B\rightarrow \rho\ell\nu$ decays and the description of inclusive $B\rightarrow X_u \ell\nu$ decays as a whole.
It opens the door to model-independent studies that will improve our understanding of fundamental parameters and light-meson spectroscopy.

\section*{Acknowledgments}
We thank Ruth Van de Water for collaboration and discussion during the initial stages of this project, Peter Stoffer for providing the P-wave phase shift of Ref.~\cite{Colangelo:2018mtw}, Alexander Khodjamirian for discussions regarding the LCSR calculations of Refs.~\cite{Hambrock:2015aor,Cheng:2017sfk,Cheng:2017smj,Descotes-Genon:2019bud,Descotes-Genon:2023ukb}, Luka Leskovec for insights into the ongoing LQCD efforts on the $B\rightarrow \pi\pi\ell\nu$ P-wave form factors, and Pablo Goldenzweig for reading the manuscript.
Furthermore, we thank Méril Reboud and Danny van Dyk for the invaluable assistance in implementing our form factor parameterization in EOS and performing the fits.
We are grateful to Michel De Cian and Veronica Kirsebom for their great interest in our work and the useful discussions regarding the upcoming LHCb measurement of $B\rightarrow \rho^0\ell\nu$ decays.
Finally, we thank Moritz Bauer for bringing the problem of the $\rho$--$\omega$ interference in the modeling of $B\rightarrow \pi\pi\ell\nu$ decays to our attention.

This research is supported in part by the Swiss
National Science Foundation (SNF) under contract 200021-212729. RvT is supported by the German Research Foundation (DFG) Walter-Benjamin Grant No. 545582477.

\begin{figure}
    \centering
    \includegraphics[width=\linewidth]{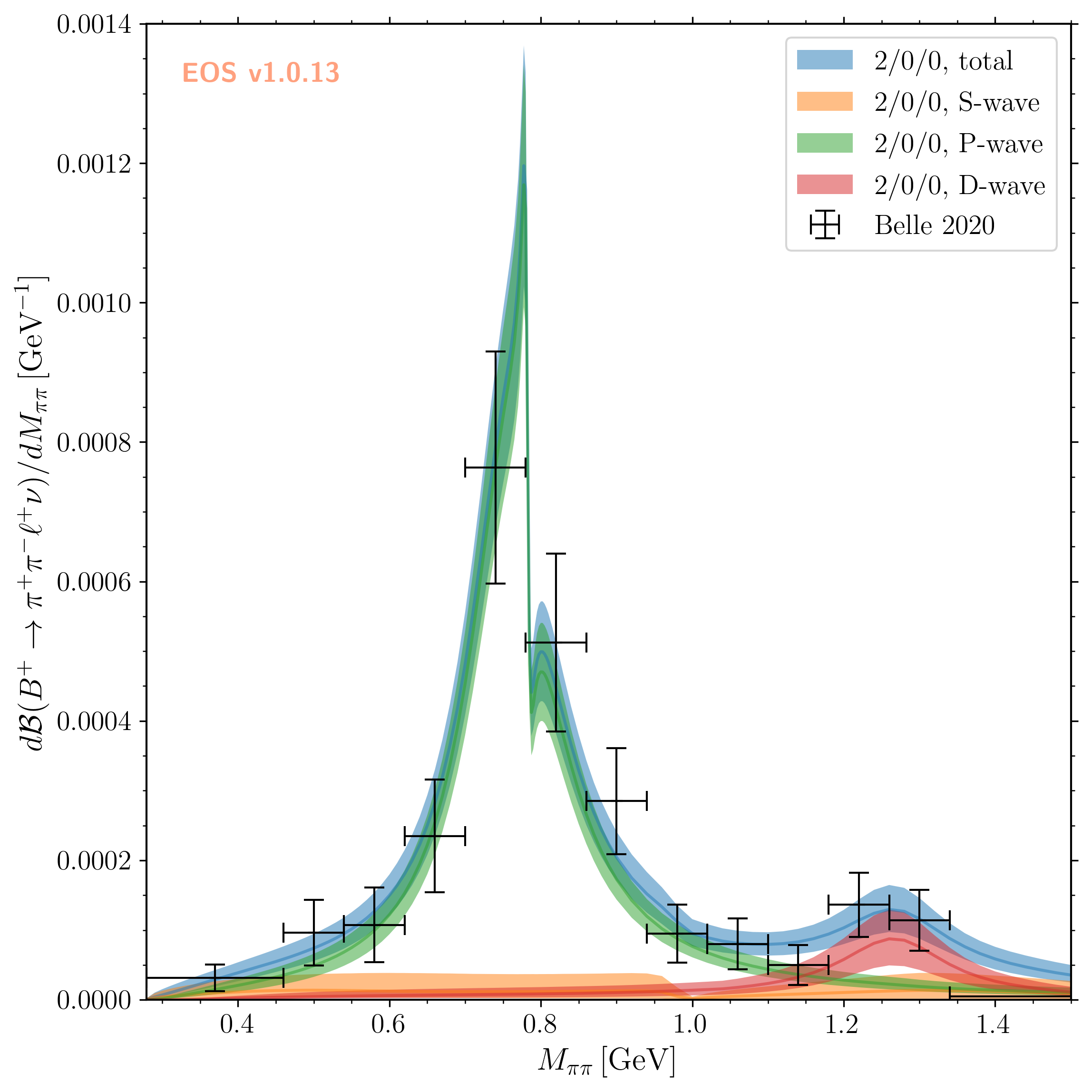}
    \caption{The $B^+\rightarrow \pi^+ \pi^- \ell^+ \nu$ $M_{\pi\pi}$ spectrum in the $2/0/0$ scenario.}
    \label{fig:mpipi_charged_200}
\end{figure}

\section*{Data availability}
EOS analysis files to reproduce the data that supports findings of this work are available at Ref.~\cite{herren_2025_15783589} and require EOS version 1.0.16 \cite{danny_van_dyk_2025_15783309}.

\appendix
\section{Further fit scenarios}\label{sec::further_fits}
In this appendix we present the $M_{\pi\pi}$ and $q^2$ spectra for lower truncation orders.

\subsection{\texorpdfstring{$2/0/0$}{2/0/0}}
The first scenario does not include any terms in the $y$-expansion and thus the lineshapes are entirely fixed by the respective Omnès functions. In comparison to the $3/2/1$ scenario in Fig.~\ref{fig:mpipi_charged}, the $2/0/0$ $M_{\pi\pi}$ spectrum in Fig.~\ref{fig:mpipi_charged_200} shows only minor differences. The most important one is the smaller uncertainty of the P-wave near the inelastic threshold, due to the absence of terms in the $y$-expansion.

\begin{figure}
    \centering
    \includegraphics[width=\linewidth]{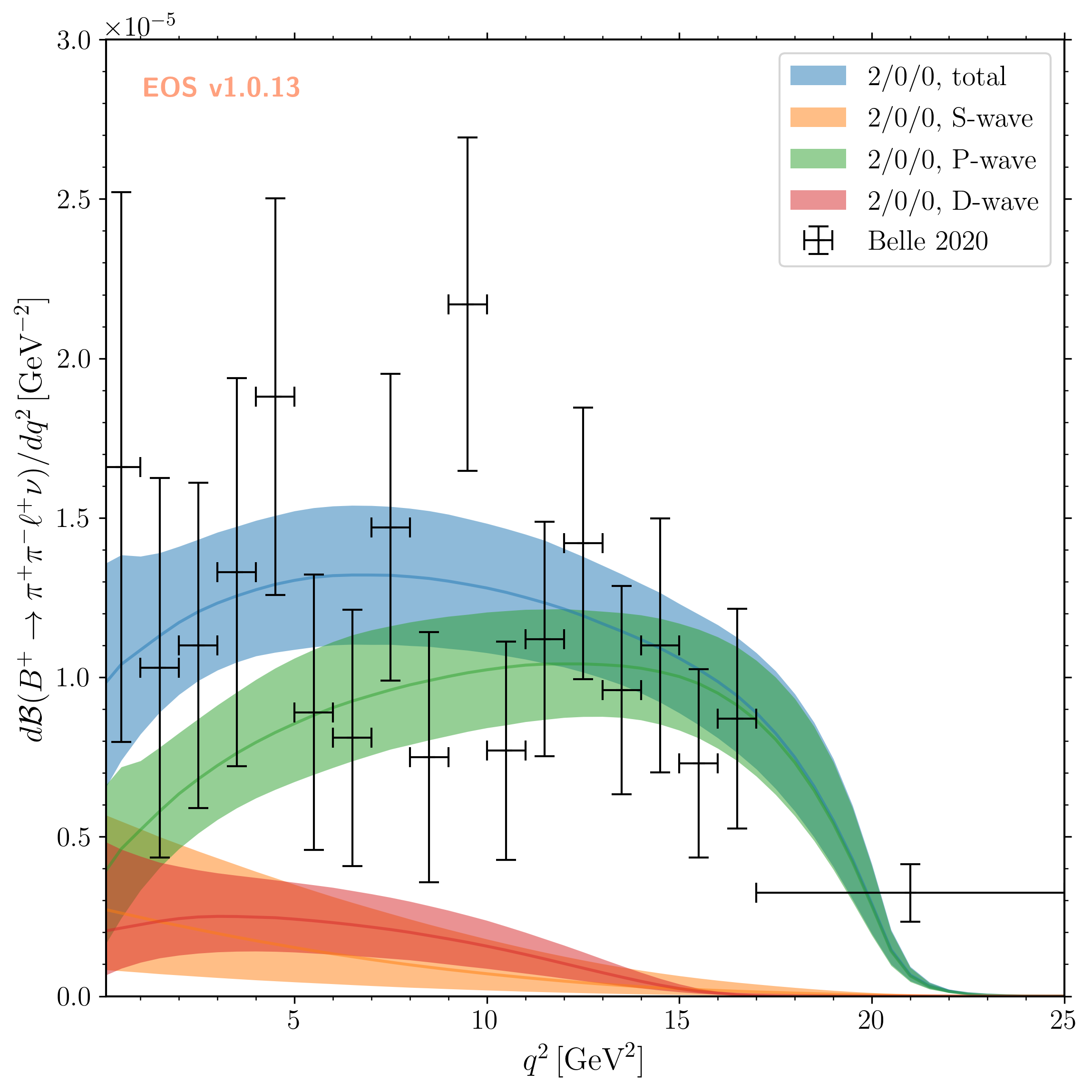}
    \caption{The $B^+\rightarrow \pi^+ \pi^- \ell^+ \nu$ $q^2$ spectrum in the $2/0/0$ scenario.}
    \label{fig:q2_charged_200}
\end{figure}

The major difference in the $q^2$ spectrum shown in Fig.~\ref{fig:q2_charged_200} is the faster drop-off of the full spectrum towards lower values of $q^2$ compared to Fig.~\ref{fig:q2_charged}. This is driven primarily by a flatter slope of the S-wave $q^2$ spectrum.

\subsection{\texorpdfstring{$2/1/0$}{2/1/0}}
\begin{figure}
    \centering
    \includegraphics[width=\linewidth]{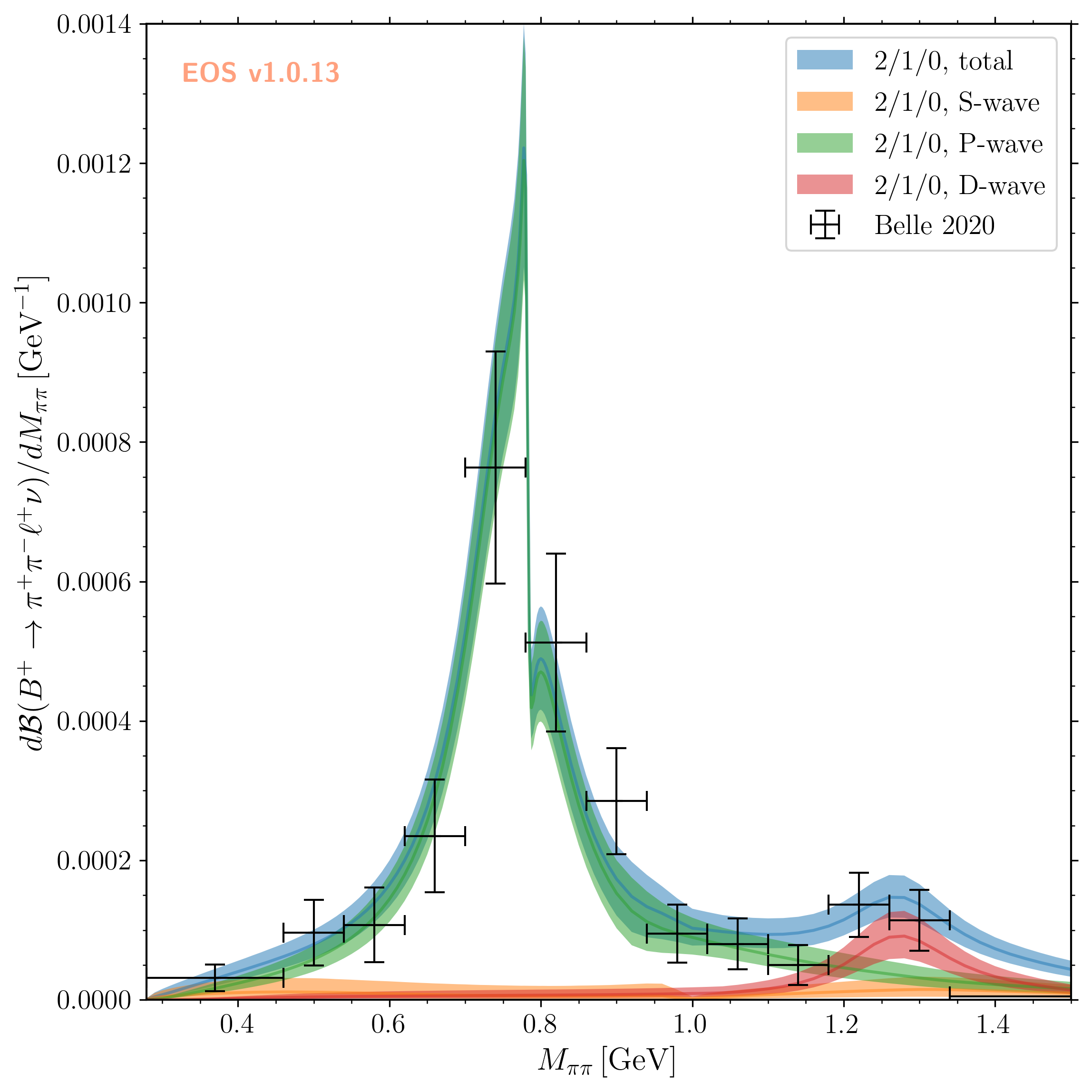}
    \caption{The $B^+\rightarrow \pi^+ \pi^- \ell^+ \nu$ $M_{\pi\pi}$ spectrum in the $2/1/0$ scenario.}
    \label{fig:mpipi_charged_210}
\end{figure}

Adding a term in the $y$-expansion leads to an increase in the uncertainty of the P-wave contribution in the $M_{\pi\pi}$ spectrum, shown in Fig.~\ref{fig:mpipi_charged_210}, around the inelastic threshold, similar to the $3/2/1$ scenario, but not quite as pronounced. Compared to the $2/0/0$ scenario, the S-wave contribution is reduced, with a slight increase of the P-wave contribution.

\begin{figure}
    \centering
    \includegraphics[width=\linewidth]{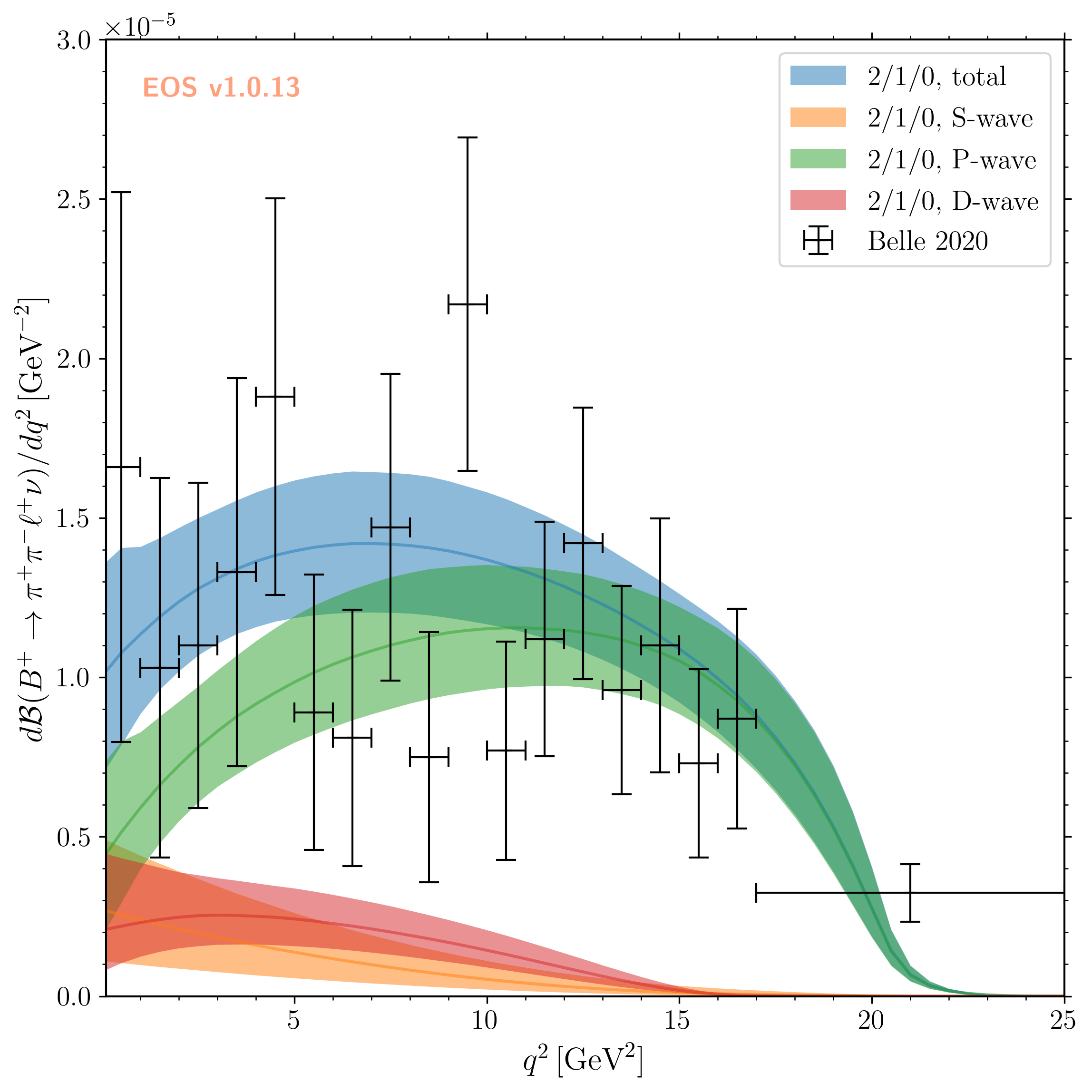}
    \caption{The $B^+\rightarrow \pi^+ \pi^- \ell^+ \nu$ $q^2$ spectrum in the $2/1/0$ scenario.}
    \label{fig:q2_charged_210}
\end{figure}

The $q^2$ spectrum shown in Fig~\ref{fig:q2_charged_210} remains unchanged with respect to the $2/0/0$ scenario which can be traced back to a similar saturation of the unitarity bounds, displayed in Figs.~\ref{fig:sat_1m} and \ref{fig:sat_1p}. 

\newpage
\bibliography{refs}

\end{document}